\documentclass[11pt]{article}
\usepackage{amsmath,amsfonts,amssymb}
\usepackage{graphicx}
\usepackage{authblk}
\usepackage{subcaption}
\usepackage{cite}
\usepackage{hyperref}
\usepackage{booktabs}
\usepackage[margin=1in]{geometry}

\title{\vspace{-2cm}Nitrogen overgrowth as a catalytic mechanism during diamond chemical vapour deposition}
 \author{ Lachlan M. Oberg$^1$, Marietta Batzer$^2$, Alastair Stacey${}^3$, and Marcus W. Doherty${}^{1,4}$}
\date{%
    $^1$Laser Physics Center, Research School of Physics, Australian National University, Australian Capital Territory 2601, Australia\\%
    $^2$Department of Physics, University of Basel, Klingelbergstrasse 82, Switzerland \\
    $^3$School of Science, RMIT University, Melbourne, Victoria 3001, Australia \\
    $^4$Quantum Brilliance Pty Ltd, 116 Daley Road, Acton, Australian Capital Territory 2601, Australia\\[2ex]%
    \today
}

\begin{document}

\maketitle

\begin{center}
    \begin{abstract}
    Nitrogen is frequently included in chemical vapour deposition feed gases to accelerate diamond growth. While there is no consensus for an atomistic mechanism of this effect, existing studies have largely focused on the role of sub-surface nitrogen and nitrogen-based adsorbates. In this work, we demonstrate the catalytic effect of surface-embedded nitrogen in nucleating new layers of (100) diamond. To do so we develop a model of nitrogen overgrowth using density functional theory. Nucleation of new layers occurs through C insertion into a C--C surface dimer. However, we find that C insertion into a C--N dimer has substantially reduced energy requirements. In particular, the rate of the key dimer ring-opening and closing mechanism is increased 400-fold in the presence of nitrogen. Full incorporation of the substitutional nitrogen defect is then facilitated through charge transfer of an electron from the nitrogen lone pair to charge acceptors on the surface. This work provides a compelling mechanism for the role of surface-embedded nitrogen in enhancing (100) diamond growth through the nucleation of new layers. Furthermore, it demonstrates a pathway for substitutional nitrogen formation during chemical vapour deposition which can be extended to study the creation of technologically relevant nitrogen-based defects.
	\end{abstract}
\end{center}

\section*{Introduction}

It is well established that trace amounts of nitrogen (N) accelerate the rate of diamond growth during chemical vapour deposition (CVD)\cite{Locher1994,Jin1994,Samlenski1995a,Ashfold2020}. While the magnitude of this enhancement varies with plasma composition, pressure, temperature\cite{Achard2007}, and activation technique\cite{Bohr1996}, a ten-fold increase in the growth rate of (100) diamond can be achieved at high plasma intensities\cite{Achard2007}. Despite the widespread adoption of N-enhanced diamond growth for both academic and commercial purposes, there does not exist a consistent and universally accepted mechanism for the effect. Moreover, a process detailing overgrowth of N on the (100) surface to form a bulk substitutional defect remains unknown. This remains a critical pre-cursor for understanding and enhancing the formation of technologically relevant N-based defects in diamond during growth. In particular, increasing the low yield and alignment of nitrogen-vacancy (NV) centers which has applications in quantum computing\cite{Taminiau2014}, communications\cite{Wehnereaam9288}, and metrology\cite{PhysRevB.88.245301,Doherty2013}.

\begin{figure*}[]
	\centering
	\includegraphics[width=1\textwidth]{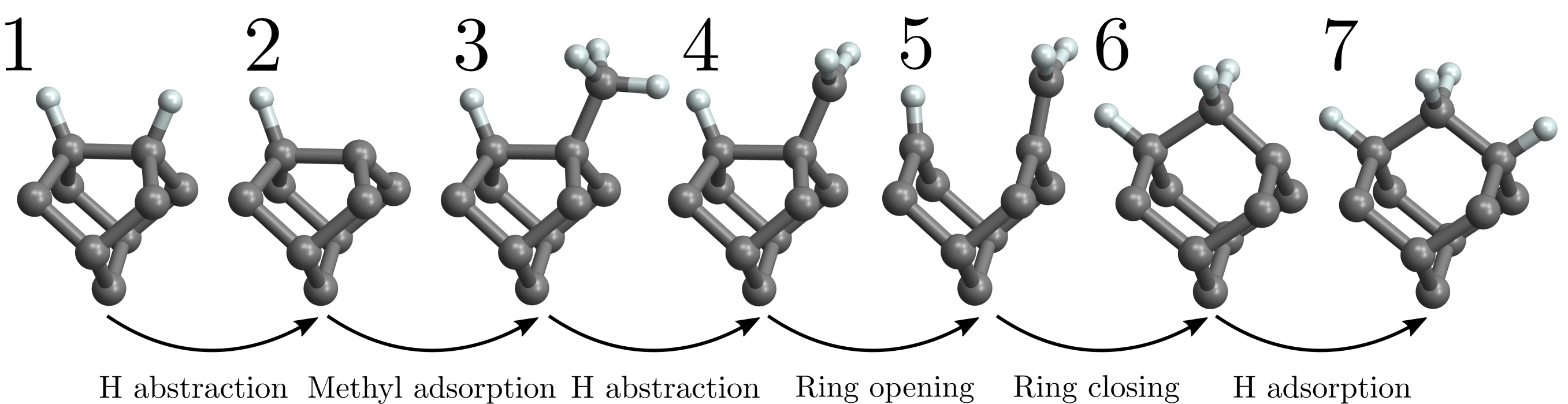}
	\caption{Conventional mechanism for C insertion into a C--C dimer of the H-terminated (100) diamond surface\cite{GARRISON835,Kang2000,Cheesman2008a}, a pivotal stage of CVD diamond growth. For simplicity we depict only the C--C dimer and neighbouring ions. Full depictions of the surface are presented in the supplementary material. In step $1\rightarrow2$, a terminating H atom is abstracted by an atomic H radical within the plasma. This creates a radical site allowing for the adsorption of a methyl radical in step $2\rightarrow3$. Further H abstraction of the methyl adsorbate ($3\rightarrow4$) creates a radical CH${}_2$ adsorbate. This adsorbate incorporates into the diamond surface via a ring opening/closing mechanism in which the dimer expands ($4\rightarrow5$) to allow formation of an energetically favourable six-membered ring ($5\rightarrow6$). Finally, the remaining surface radical is stabilized the adsorption of atomic H in step $6\rightarrow7$.}
	\label{fig:mechanism}
\end{figure*}

While the catalytic role of N has generated considerable theoretical interest, existing literature has failed to identify an atomic mechanism which can demonstrably produce order-of-magnitude enhancements to diamond growth rates. The majority of studies have focused on the role of sub-surface N and N adsorbates in reducing reaction and transition state energies for hydrogen-terminated (100) diamond growth\cite{Frauenheim1998,Kaukonen1998,VanRegemorter2008,VanRegemorter2009,VanRegemorter2009a,Yiming2014b}. These \textit{ab initio} works have primarily studied the impact of N on carbon (C) insertion into a C--C dimer of the reconstructed (100) surface\cite{Kang2000,Cheesman2006,Cheesman2008a,Oleinik2000,Tamura2005}. This process has received substantial attention because it is fundamental for all (100) diamond growth by CVD. Most studies follow or slightly develop the widely-accepted model for C insertion presented by Garrison~\textit{et al.}\cite{GARRISON835}. This plasma-surface reaction is itself built on extensive earlier work\cite{Harris1990a,Frenklach1991,Huang1988,Frenklach1988} and consists of six key steps depicted in Figure~\ref{fig:mechanism}.

Early work by Frauenheim~\textit{et al.} demonstrated that sub-surface N reduced the binding energy of terminating H atoms and therefore enhanced the rate of initial H abstraction\cite{Frauenheim1998}. This was attributed to charge stabilization of the resulting surface radical (structure 2 in Figure~\ref{fig:mechanism}) due to electron transfer from the substitutitonal N lone pair\cite{Samlenski1996}. However, this makes methyl adsorption (step $2\rightarrow3$) energetically unfavourable. An alternative growth mechanism was therefore proposed involving charge transfer to the metastable anti-bonding orbital of C--C surface dimers and subsequent adsorption of CH${}_2$ radicals. This hypothesis is challenged by CVD plasma experiment and modeling which demonstrated that methyl radicals (CH${}_3$) are the majority C-based radical species proximate to the diamond surface during growth and exceed CH${}_2$ concentrations by three orders of magnitude\cite{May2007a,Mankelevich2008c}.

Alternatively, more recent works have rationalised Frauenheim \textit{et al.}'s original findings by assuming that H abstraction, not methyl adsorption, is the rate-limiting step for diamond growth. These studies identify that near-surface N defects enhance H abstraction from methyl adsorbates ($3\rightarrow4$) through charge transfer\cite{VanRegemorter2008}. Additionally, near-surface N defects promote the creation of adsorption sites ($1\rightarrow2$) through weakening of proximal C--C surface dimer bonds\cite{Yiming2014b}. Further \textit{ab initio} work has demonstrated that sub-surface N increases activation barriers for CH${}_2$ insertion into the C--C dimer and CH${}_2$ surface migration\cite{VanRegemorter2009a}. Other studies have found that co-adsorbed N atoms have no bearing on H abstraction\cite{Yiming2014b}, reduces the adsorption rate of CH${}_3$ on step edges, but increases the rate of adsorption (and hypothesized to enhance migration) of CH${}_2$ on step-edges\cite{VanRegemorter2009}. It has also been proposed that N accelerates (111) growth by nucleating new diamond layers\cite{Butler2008}. In summary, extensive \textit{ab initio} modeling has not yet provided a clear consensus on the catalytic role of N in CVD.

As a supplement to \textit{ab initio} calculations, extensive Monte-Carlo modeling has been performed to understand the complexity of mesoscale diamond growth. These works have highlighted the importance of surface migration\cite{NETTO20051630} and nucleation of new layers\cite{Battaile1997} to reproduce experimental growth rates and morphologies. For example, the surface of CVD-grown (100) diamond often exhibits a relatively smooth and terraced structure\cite{Butler2009}. However, note that this is not ubiquitous and depends on reactor conditions\cite{Li2006,Achard2007}. Such morphologies are indicative of step-flow modes, in which migration of hydrocarbon adsorbates and preferential adsorption at step-edges is believed to contribute largely to layer growth. Recent Monte-Carlo modeling has emphasized the importance of critical nuclei for propagating layer growth. These are immobile surface features, such as a lone C--C dimer, which act as nucleation points\cite{May2009,Rodgers2015}. Furthermore, the inclusion of super-nucleating species in Monte-Carlo models has been found to catalyze diamond growth. These are adsorbates or surface defects (hypothesised to be N-based) which quickly form critical nuclei following formation. When growth is limited by nucleation of new layers (i.e., growth is dominated by step-flow), super-nucleating species have demonstrated ten-fold enhanced growth rates\cite{May2009}. The search for a N-based super-nucleating species is therefore well founded and forms the primary aim of this work.

\begin{figure}[]
  \centering
  \includegraphics[width=0.3\textwidth]{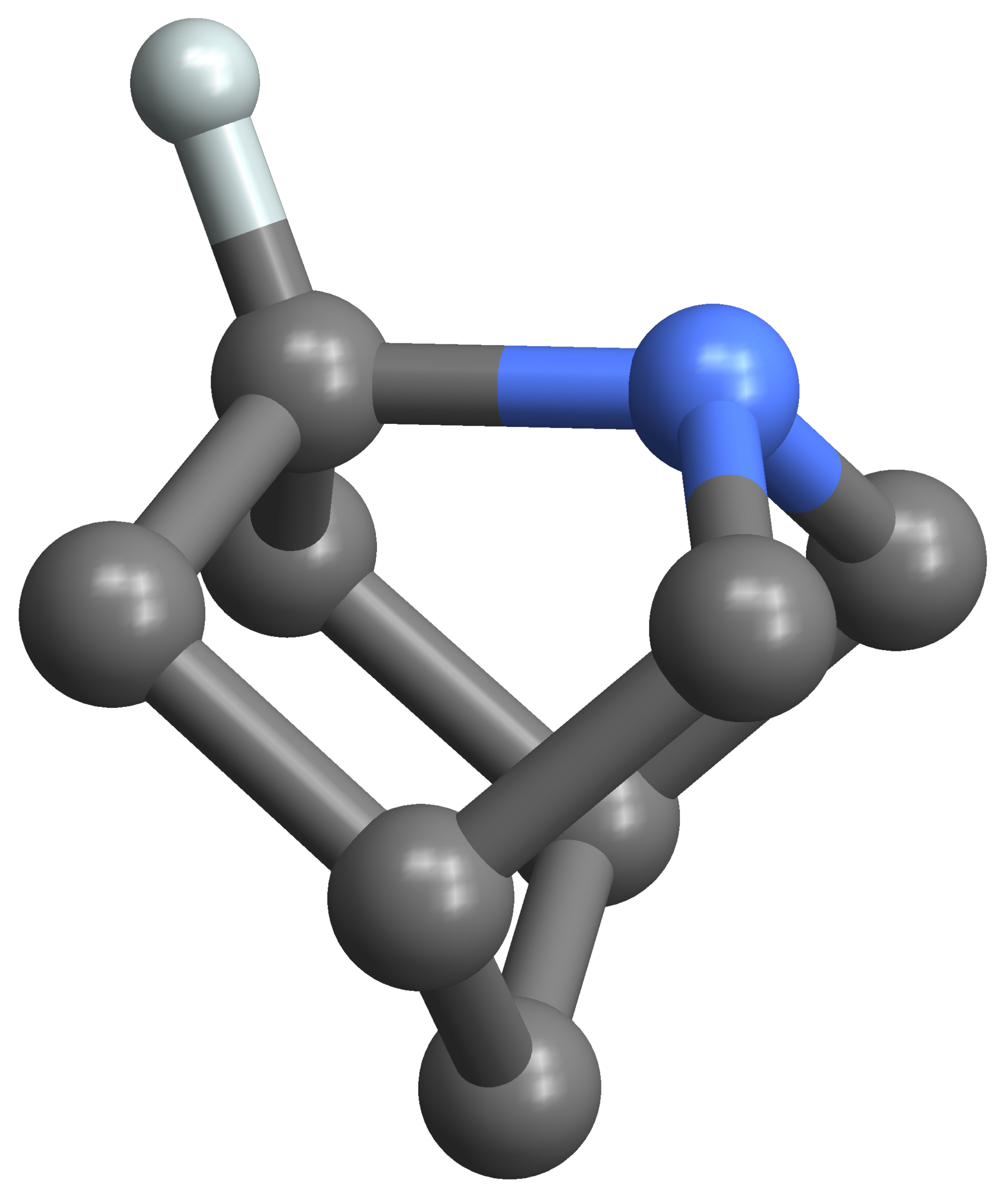}
  \caption{Surface-embedded N on the (100) surface forms a dimer unit with a neighbouring C atom.}
  \label{fig:C--N}
\end{figure}

We identify a potential super-nucleating species by producing the first atomistic model for sub-surface N formation during CVD. Our calculations demonstrate that surface-embedded N catalyzes the nucleation of new diamond layers during its overgrowth and subsequent encapsulation into bulk diamond. The atomic structure of surface-embedded N is depicted in Figure~\ref{fig:C--N}. It consists of a substitutional defect which maintains the structure of the dimerised 2x1-(100) surface. The bonding of the C--N dimer resembles that of a typical C--C surface dimer with a C-H covalent bond substituted by a N electron lone pair. Surface-embedded N has previously been identified as the most energetically stable form of substitutional N in the (100) diamond surface\cite{PhysRevB.88.245301}. Furthermore, it is reported to be evident in NEXAFS scans of diamond (100) surfaces following N plasma treatment\cite{doi:10.1002/admi.201500079}. Consequently, it is likely that surface-embedded N is a common surface defect during CVD growth of the (100) surface. It is therefore chosen as the initial point for our overgrowth mechanism.

Further motivation for this work includes the refinement of existing \textit{ab initio} techniques used in previous CVD-growth literature. Firstly, many works employ cluster models\cite{VanRegemorter2009a,Kang2000,Tamura2005,Oleinik2000,Cheesman2008a} to represent the diamond surface. While cluster models are capable of reproducing results consistent with periodic slab calculations\cite{Steckel2001,Tracey2013}, this is not always the case\cite{Nigam2014}, and careful optimisation of cluster dimensions is always required. As has been noted in previous works\cite{Cheesman2008a}, the relaxed steric constraints associated with some cluster calculations result in an underestimation of structural stabilities during the ring opening/closing mechanism\cite{Oleinik2000}. Secondly, some diamond studies\cite{Cheesman2008a,Yiming2014b} employ the drag method for determining the transition state (or do not explicitly mention the method used\cite{Kang2000,Tamura2005,Oleinik2000}) which can produce inaccurate activation barriers for complex potential energy surfaces\cite{Halgren1977,Xie2004}. Instead, it is preferable to use chain-of-states methods when possible, such as the nudged-elastic-band (NEB) technique, which provide greater consistency in determining the transition state\cite{Henkelman2002}. Consequently, in this work we employ state-of-the-art density functional theory, including a fully quantum slab model of the H-terminated diamond (100) surface, the use of hybrid functionals, and the NEB method for determining transition states.

Our model for overgrowth begins with surface-embedded N as presented in Figure~\ref{fig:C--N} and is separated into three distinct stages. Firstly, nucleation of a new diamond layer through C insertion into the C--N dimer is presented in Section~\ref{sec:I}. We find that surface-embedded N reduces energy requirements for nearly every stage of the reaction depicted in Figure~\ref{fig:mechanism}. In particular, the rate of the ring opening/closing mechanism ($4\rightarrow5\rightarrow6$) is increased by a factor of approximately 400. This provides a compelling atomistic mechanism for the N catalysis of diamond CVD growth through enhanced layer nucleation. In Section~\ref{sec:II} we present the second growth stage where we consider layer growth emanating from the nucleation point. We demonstrate the formation of a new C--C dimer and C bridging to form a new dimer row. The key result is that surface-embedded N does not impede typical diamond layer growth. The final growth stage is considered in Section~\ref{sec:III}, where we demonstrate that full encapsulation of the N defect into the surface (i.e., the formation of sub-surface N) is mediated by charge transfer.

\section{Method}

Density functional theory was performed using the VASP plane-wave code\cite{Kresse1996a,Kresse1994a,Kresse1996,Joubert1999} using PBE\cite{Perdew1996b} and verified using B3LYP\cite{Lee1988} functionals. PBE belongs to the generalised-gradient approximation (GGA) class of exchange-correlation functionals. While GGA functionals are computationally efficient, they are known to over-delocalize electrons and over-stabilize stretched bonds of transition states. This can result in underestimations of energy barriers, especially for gas-surface reactions\cite{Janesko2015}. Hybrid functionals, such as B3LYP, correct the over-delocalisation of GGA functionals by including some contribution of exact exchange energy thereby improving estimations of reaction barriers\cite{doi:10.1021/ct100488v}. However, this increase in accuracy requires significant computational costs and so we perform the majority of calculations using the PBE functional. To validate the accuracy of these PBE calculations we calculate the key ring opening step ($4\rightarrow5$ of Figure~\ref{fig:mechanism}) using B3LYP functionals as discussed in the following section.

The $2\times1$ reconstructed H-C(100) surface has been represented using a periodic slab model consisting of 240~C ions with dimensions $5\times4\times6$ (in units of bulk-diamond primitive unit cells). As presented in the supplementary material, these dimensions have been stringently optimised against the surface work function, reaction and transition state energies for the ring opening/closing mechanism, and interaction energies between N defects in adjacent supercells. Slab images in the \textless100\textgreater \ direction were separated by a distance of 10~\AA \ and long-range dipole interactions have been corrected for. $\Gamma$-point sampling and real-space projection operators have been used for all calculations. The cut-off energy for the plane-wave basis is 600~eV. The electronic tolerance for successive iterations of the self-consistent field method is taken as $0.1$~meV, while all geometry optimisations are performed to a tolerance of $0.05$~eV/\AA \ per ion. Ions composing the top two layers of primitive unit cells (those primarily involved in surface reactions) were allowed to relax in all directions, while the remaining ions were permitted to relax only in the \textless100\textgreater \ direction. The effects of spin polarization on both stable and transition states has been fully accounted for. The climbing NEB technique has been used to calculate transition states for all reaction barriers\cite{doi:10.1063/1.1329672}. This extension to NEB determines the exact transition state by forcing the highest energy image to the saddle point.

\section{Nucleation of a new diamond layer}
\label{sec:I}

The first stage of growth consists of new layer nucleation through C insertion into the C--N dimer. The results of our \textit{ab initio} calculations are presented in Table~\ref{tab:results}, which documents the reaction energies ($\Delta E$) and transition state energies (TS) for each of the steps required for C insertion as per Figure~\ref{fig:mechanism}. This is performed using PBE functionals for both C--C and C--N dimers for the purposes of comparison. We also include several B3LYP calculations performed using both our slab geometry as well as a small adamantane cluster (C${}_{10}$H${}_{16}$) as discussed below. Additionally, we compare our results to previous values for C--C dimers by Cheesman~\textit{et al}\cite{Cheesman2008a}. This is a large-scale hybrid study in which the reactive C--C dimer and neighbouring atoms were treated using B3LYP functionals, while the remaining $\sim1500$ atoms of the cluster were treated using molecular mechanics (MM). We have chosen this particular work as it reproduces or demonstrably improves on previous studies for C insertion that employ both fully quantum\cite{Cheesman2006,Kang2000} and hybrid cluster models\cite{Oleinik2000,Tamura2005}.

\begin{table*}[]
\begin{center}
\begin{tabular}{ccccccccccc}
                               &            & \multicolumn{2}{c}{PBE}    &  & \multicolumn{2}{c}{B3LYP}  &  & B3LYP        &  & B3LYP                                                                            \\
                               &            & \multicolumn{2}{c}{(slab)} &  & \multicolumn{2}{c}{(slab)} &  & (adamantane) &  & (Cheesman \textit{et al.}\cite{Cheesman2008a}) \\ \cline{3-11} 
\textbf{Step} &            & C--C         & C--N        &  & C--C         & C--N        &  & C--C         &  & C--C                                                                             \\ \hline
$1\rightarrow2$                & $\Delta E$ & -0.026       & -0.135      &  &              &             &  &              &  & -0.007                                                                           \\
                               & TS         & 0.185        & 0.136       &  &              &             &  &              &  & 0.274                                                                            \\ \hline
$2\rightarrow3$                & $\Delta E$ & -4.556       & -3.671      &  &              &             &  &              &  & -3.868                                                                           \\
                               & TS         & 0            & 0           &  &              &             &  &              &  & 0                                                                                \\ \hline
$3\rightarrow4$                & $\Delta E$ & -0.228       & -0.142      &  &              &             &  &              &  & -0.312                                                                           \\
                               & TS         & 0.355        & 0.358       &  &              &             &  &              &  & 0.306                                                                            \\ \hline
$4\rightarrow5$                & $\Delta E$ &              & -0.143      &  &         &   -0.235        &  & 0.019        &  & 0.290                                                                            \\
                               & TS         &              & 0.057       &    &        &  0.094          &  & 0.370        &  & 0.450                                                                            \\ \hline
$5\rightarrow6$                & $\Delta E$ &              & -0.476      &  &              &   -0.329      &  & -0.540       &  & -0.813                                                                           \\
                               & TS         &              & 0.283       &  &              &             &  & 0.571        &  & 0.550                                                                            \\ \hline
$4\rightarrow6$                & $\Delta E$ & -0.350       &             &  &     -0.299         &               &  &              &  & -0.513$^\ddag$                                     \\
                               & TS         & 0.959        &             &  &    0.965${}^\dag$        &         &  &              &  & $0.900^\ddag$                                      \\ \hline
$6\rightarrow7$                & $\Delta E$ & -4.141       & -4.166      &  &              &             &  &              &  & -4.573                                                                           \\
                               & TS         & 0            & 0           &  &              &             &  &              &  & 0                                                                               
\end{tabular}
\end{center}
\caption{Reaction ($\Delta E$) and transition state (TS) energies for C insertion into C--C or C--N dimers on the H-C(100) surface (c.f. Figures~\ref{fig:mechanism} and~\ref{fig:Nharris}). All energies are in eV. ${}^\dag$Transition state geometry optimised using 400~eV cut-off energy. ${}^\ddag$Effective energy for the direct transition $4\rightarrow6$ if state 5 was not stable.}
\label{tab:results}
\end{table*}

\begin{figure*}[]
  \centering
  \includegraphics[width=1\textwidth]{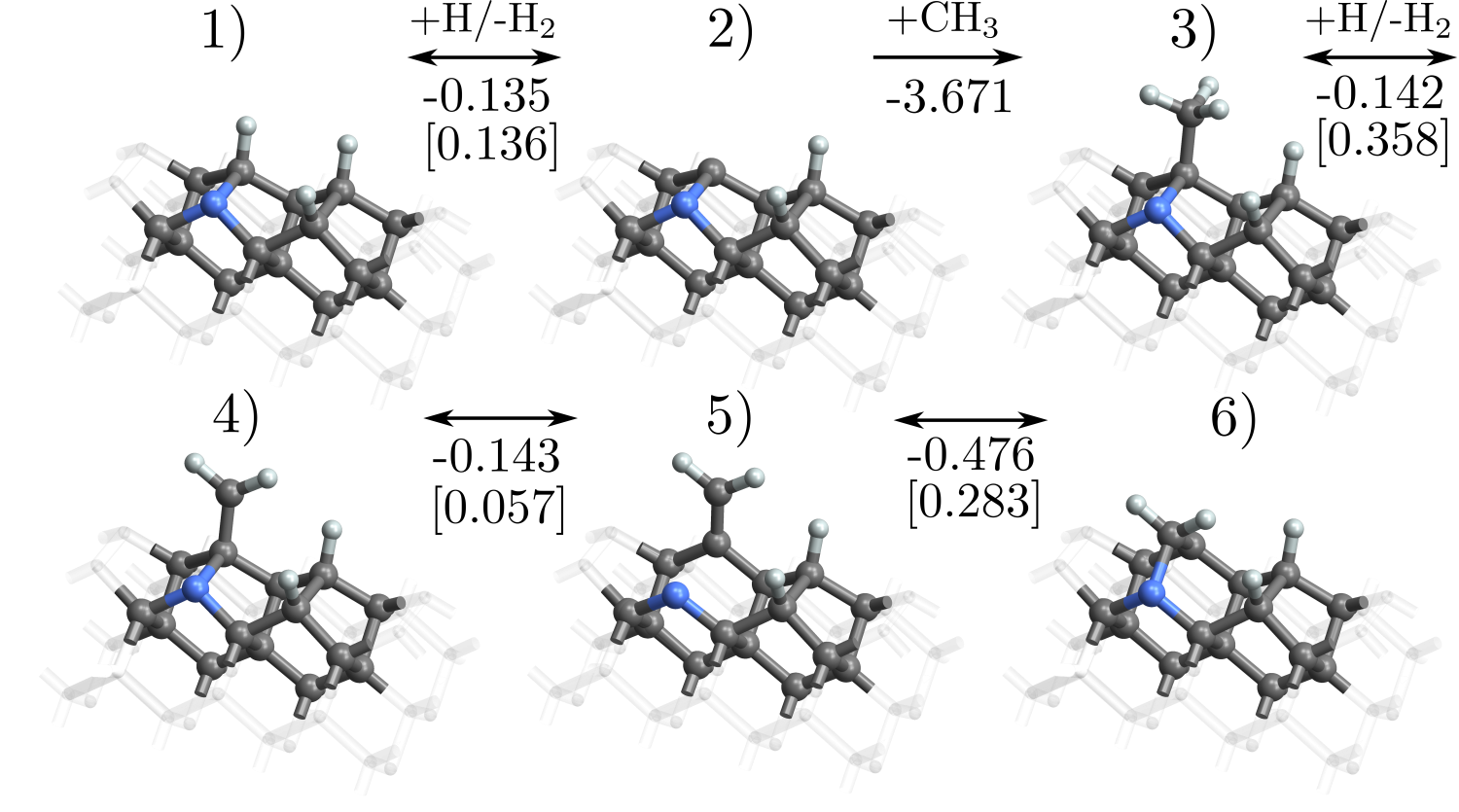}
  \caption{The mechanism for C insertion into a C--N dimer on the C(100) surface. Reaction energies (in eV) are presented between each step in the reaction, and where applicable we also include barrier energies in square brackets.}
  \label{fig:Nharris}
\end{figure*}

\begin{figure}[]
  \centering
  \includegraphics[width=0.3\textwidth]{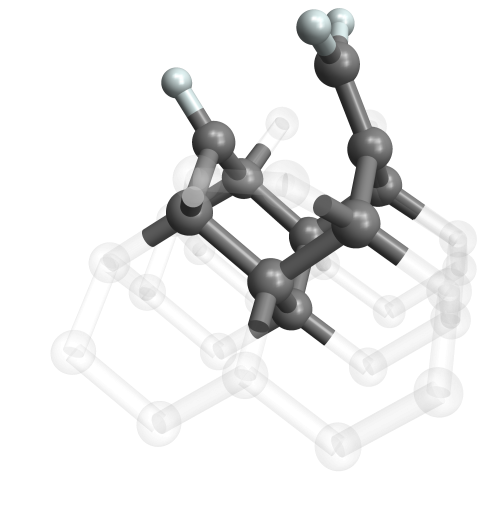}
  \caption{Transition state for the ring opening/closing reaction ($4\rightarrow6$) during C insertion into a C--C dimer.}
  \label{fig:TS46}
\end{figure}

\subsection{C insertion into C--C}

Our results for C insertion into the C--C dimer are largely consistent with existing \textit{ab initio} works. We identify energy barriers and transition states for initial H abstraction ($1\rightarrow2$) similar to Cheesman \textit{et al.} and several other studies\cite{Kang2000,Cheesman2006}. However as expected the PBE functional underestimates the transition state energy by approximately 100~meV. As with other works, we identify no barrier to CH${}_3$ adsorption ($2\rightarrow3$) or H adsorption ($6\rightarrow7$)\cite{Cheesman2006,Cheesman2008a,Tamura2005,Oleinik2000}. While our transition state energy for the H abstraction in step $3\rightarrow4$ overestimates that of Cheesman \textit{et al.} by 49~meV, it underestimates that obtained in the B3LYP cluster studies of Kang and Musgrave by approximately 100~meV\cite{Kang2000}.

The greatest deviation between our study and existing literature is for the ring opening/closing mechanism, $4\rightarrow5\rightarrow6$. We now discuss this discrepancy in-depth due to its relevance for the validity of our results and to a general understanding of conventional diamond growth. The majority of studies find that state 5, the open-ring configuration, is stable with the CH${}_2$ adsorbate realizing an sp${}^2$ bonding configuration (c.f. Figure~\ref{fig:mechanism})\cite{Cheesman2008a,Cheesman2006,Tamura2005,Kang2000}. However, in agreement with the GGA cluster calculations of Oleinik~\textit{et al.}\cite{Oleinik2000}, we find that state 5 is not stable and that C insertion occurs through the direct transition $4\rightarrow6$. The transition state for $4\rightarrow6$ does not possess the sp${}^2$ bonding configuration of state 5. Instead, as depicted in Figure~\ref{fig:TS46}, the CH${}_2$ radical inserts into the ring structure simultaneously with breakage of the dimer bond.

We have confirmed that the instability of state 5 is not a deficiency of the PBE functional. B3LYP functionals in combination with a range of geometry optimisation algorithms (RMM-DIIS\cite{PULAY1980393}, conjugate gradient, and velocity damping) also failed to identify a stable sp${}^2$-like intermediate between states 4 and 6. As presented in Table~\ref{tab:results}, we also perform climbing NEB calculations with B3LYP for the direct transition $4\rightarrow6$ and obtain a transition state consistent with the PBE functional. Note that due to computational constraints the transition state geometry was optimised using a lower plane wave cut-off energy of 400~eV. However, as demonstrated in Figure~1 of the supplementary material, this does not significantly reduce the accuracy of our calculations.

Instead, the stability of state 5 appears to be conditional on the use of either cluster or hybrid quantum/MM models. We have calculated the energetics for ring opening/closing using B3LYP functionals on an adamantane molecule (C${}_{10}$H${}_{16}$, resembling the models in Figure~\ref{fig:mechanism} if all dangling bonds were hydrogen terminated), the simplest possible cluster model for a surface dimer. Note that the corresponding C${}_{10}$ carbon cluster is also the fully quantum region used in several hybrid studies\cite{Cheesman2008a,Oleinik2000,Tamura2005}. Here we identify that state 5 is stable and that the reaction $4\rightarrow5\rightarrow6$ possesses transition state and reaction energies in close agreement with Cheesman~\textit{et al.} as shown in Table~\ref{tab:results}. Moreover, we find that the geometry of state 5 in our cluster calculations is considerably relaxed with respect to our slab calculations. In particular, the separation distance between C atoms of the broken dimer of state 5 is large at 2.95 \AA. As demonstrated in the supplementary, more than 1.4~eV is required to achieve a separation distance of this scale within the slab geometry. It is clear that the periodicity enforced by slab models limits the relaxation of the reactive dimer and surrounding ions. Clusters are therefore inadequate for modeling some aspects of diamond surface chemistry which are better suited to periodic slabs.

Without the erroneous identification of state 5 as stable, cluster models produce results in agreement with our slab model. To see this, consider the effective energies for the direct transition $4\rightarrow6$ deduced from the work of Cheesman~\textit{et al}. As shown in Table~\ref{tab:results}, if state 5 is ignored we obtain a transition-state energy of 0.900~eV and a reaction energy of -0.513~eV, similar to respective values of 0.959~eV and -0.350~eV for our work using PBE functionals.

\subsection{C insertion into C--N}

The presence of surface-embedded N greatly enhances C insertion. Firstly, our results in Table~\ref{tab:results} show that the transition barrier for the initial H abstraction (step $1\rightarrow2$) is reduced by 49~meV while stability is increased by 109~meV. This is attributable to the greater electronegativity of N in comparison to C, where we note that the length of the C--N dimer bond is reduced by 5.5\% with respect to C--C. A similar reduction in the transition state energy is not observed for the abstraction $3\rightarrow4$ as this reaction step does not directly involve bonding with surface-embedded N.

The greatest enhancement to C insertion due to surface-embedded N occurs during the ring opening/closing mechanism. In contrast to the C--C dimer, state 5 for the C--N dimer forms a stable but shallow minimum in which the CH${}_2$ adsorbate realizes an sp${}^2$ bonding configuration. Moreover, as per our PBE calculations the transition $4\rightarrow5$ requires an activation energy of approximately 5~meV; negligible at typical diamond growth temperatures. Given the significance of transition $4\rightarrow5$ for our model of N catalysis, we validate the accuracy of these PBE calculations using B3LYP functionals. Full geometry optimisation and the climbing NEB method produce $\Delta E = -0.235$~eV and a transition state energy of $0.094$~eV. These are in good agreement with the PBE values of $-0.143$~eV and $0.057$~eV respectively. This indicates that the PBE functional possesses similar accuracy to B3LYP for diamond surface reactions (potentially excluding gas-surface interactions as discussed above), significantly reducing computational costs.

The combination of this low transition state energy and the stability of state 5 drastically enhances the rate of the ring opening/closing mechanism. This can be quantified using kinetic rate equations to compare the reaction times between C--C and C--N dimers. Defining the population of states 4, 5, and 6 as $\phi_4$, $\phi_5$, and $\phi_6$ respectively we have that
\begin{align}
\phi_4'(t) &= -\Gamma_{4\rightarrow5}\phi_4 + \Gamma_{5\rightarrow4}\phi_5 \label{system}\\
\phi_5'(t) &= -(\Gamma_{5\rightarrow4}+\Gamma_{5\rightarrow6})\phi_5 + \Gamma_{4\rightarrow5}\phi_4 + \Gamma_{6\rightarrow5}\phi_6 \nonumber \\
\phi_6'(t) &=-\Gamma_{6\rightarrow5}\phi_6 + \Gamma_{5\rightarrow6}\phi_5. \nonumber 
\end{align}
Here $\Gamma_{i\rightarrow j}=A_{i\rightarrow j}e^{-\text{TS}_{i\rightarrow j}/k_B T}$, where $A_{i\rightarrow j}$ is the attempt frequency for the transition between states $i$ and $j$, $\text{TS}_{i\rightarrow j}$ is the transition state energy, and $T=1200$~K is the approximate temperature of the diamond surface during CVD\cite{Butler2009}. For C insertion into the C--C dimer, one only needs to consider the direct transition $4\rightarrow6$ and the rate equations are simply
\begin{align*}
\phi_4'(t) &= -\Gamma_{4\rightarrow6}\phi_4 + \Gamma_{6\rightarrow4}\phi_6\\
\phi_6'(t) &=-\Gamma_{6\rightarrow4}\phi_6 + \Gamma_{4\rightarrow6}\phi_4.
\end{align*}

A quantitative determination of the pre-exponential factors requires calculating the vibrational partition functions of each stable and transition state\cite{Bao2017}. To compare the pre-exponential rate of ring opening/closing between C--C and C--N dimers, we may instead approximate the ratio of thermal occupation between their respective normal modes. We find that this does not differ by more than a factor of three for vibrational energies in the expected range of $100-200$~meV. Therefore for simplicity we assume that $A_{i\rightarrow j}=A$ is constant for all processes involved in ring opening and closing. This allows us to solve the above system of equations given the initial conditions $\phi_4(0)=1$ and $\phi_5(0)=\phi_6(0)=0$. Processes interrupting ring opening/closing, such as H adsorption and abstraction, are not included. These are assumed to impact C insertion into C--C and C--N at equal rates and therefore do not influence our relative comparison.

\begin{figure}[]
	\centering
	\includegraphics[width=0.4\textwidth]{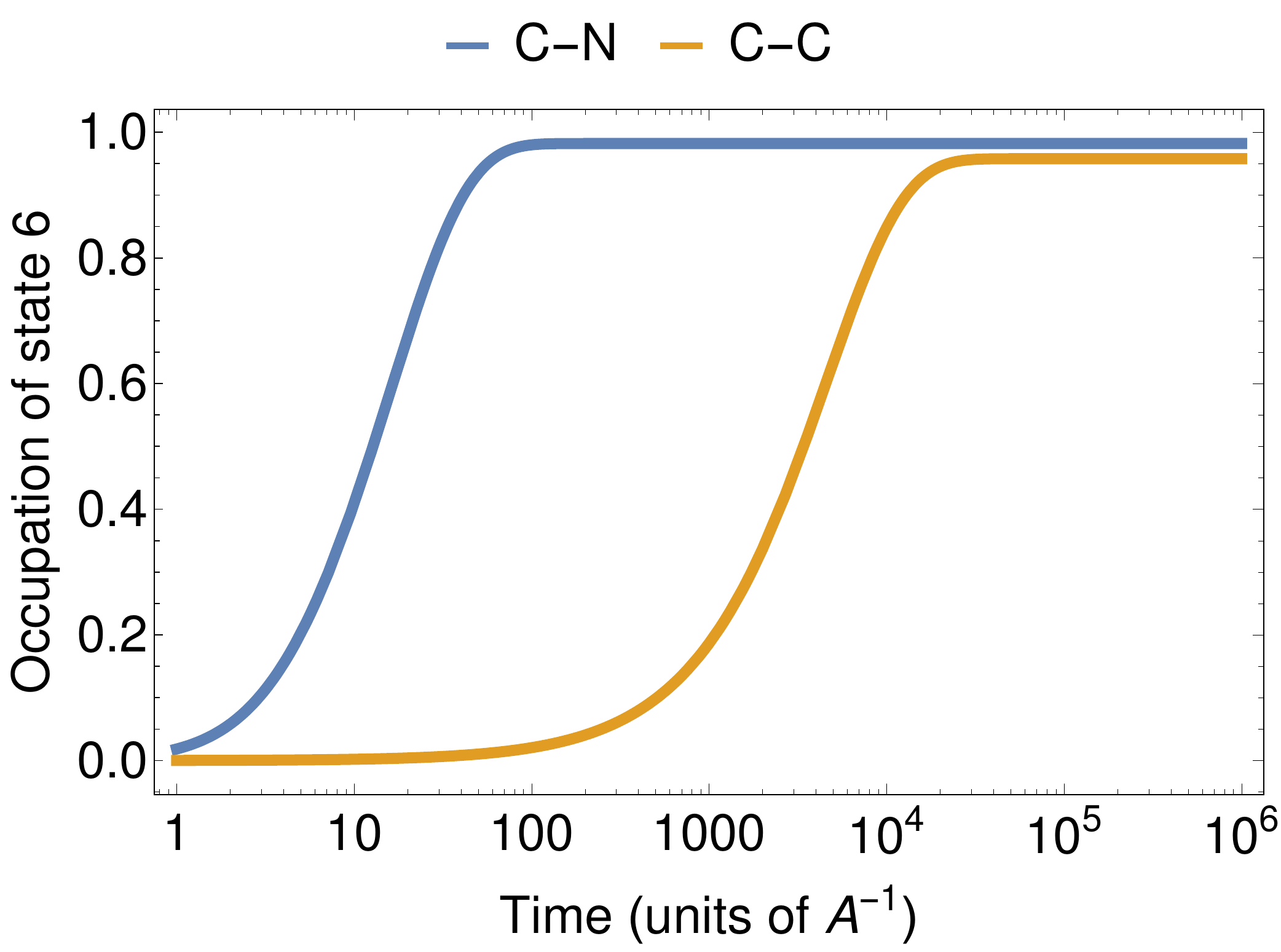}
	\caption{Occupation of state six as a function of time for the ring opening/closing mechanism on a C--C dimer (orange) and C--N dimer (blue).}
	\label{fig:CVsN}
\end{figure}

In Figure~\ref{fig:CVsN} we plot $\phi_6$ as a function of time for ring opening/closing with C--C and C--N dimers. These results indicate that ring closure occurs approximately 400 times faster in the presence of surface-embedded N. Not only does this drastically enhance the rate of C insertion, it also reduces the residence time of state 4. This would likely result in less etching of the C adsorbate, believed to occur predominately via $\beta$-scission\cite{Goodwin2019,NETTO20051630,Butler2009}.

Given the large transition state energies required for ring opening/closing during conventional diamond growth, step $4\rightarrow6$ appears to be a limiting step for new layer nucleation. Consequently, we propose that surface-embedded N can catalyze new layer growth by substantially reducing energy requirements for C insertion into the surface. This enhancement is expected to be most pronounced during (100) step-flow modes when growth rates are limited by new layer nucleation. Under such circumstances, we propose that surface-embedded N acts as a super-nucleation species for new-layer C growth and subsequently the formation of critical nuclei on the surface. This provides an atomic mechanism for enhanced growth rates observed in some Monte-Carlo studies\cite{May2009}. Note, however, that the catalytic mechanism identified here cannot be universal for all N-enhanced diamond growth, particularly that observed for diamond surfaces other than (100).

It is difficult to quantify the enhancement this catalytic effect has on macroscopic growth rates. This would require a mesoscopic study (such as a Monte-Carlo simulation) which includes the effects of surface-embedded N amongst the complex diversity of surface chemistry that occurs during CVD. However, amongst existing \textit{ab initio} studies of (100) diamond growth, this work demonstrates the greatest enhancement to any key reaction process. For example, previously suggested catalytic effects of N include enhanced H-abstraction through weakening of surface bonds proximate to sub-surface N\cite{Yiming2014b}. This increases the rate of H abstraction local to the sub-surface N by 2.4. This enhancement is minor compared to the 400-fold increase in the rate of ring opening and closing in the presence of surface-embedded N. By this metric, our work demonstrates the most compelling atomistic mechanism for N catalysis to date.

\section{Layer growth}
\label{sec:II}

In this section we demonstrate that layer growth propagating from the nucleation point is not hindered by the presence of surface-embedded N. Due to the complexity of the CVD environment there are many possible processes that can contribute to layer growth. Hence, we consider two representative mechanisms; formation of a new surface dimer and C addition between dimer rows. We note that these two mechanisms are not necessarily the atomic processes that produce step-flow growth (which have not yet been conclusively identified in existing literature). Instead, the two representative mechanisms involve key elements ubiquitous to most layer growth processes and therefore assumed to be highly relevant for step flow. These elements include H abstraction, CH${}_3$ adsorption, ring opening/closing, and migration. We calculate the reaction and transition state energies of these two mechanisms in the presence of surface-embedded N and compare these results to existing literature of analogous processes without N\cite{Cheesman2008a}.

\subsection{Dimer formation}

\begin{figure*}[]
	\centering
	\includegraphics[width=1\textwidth]{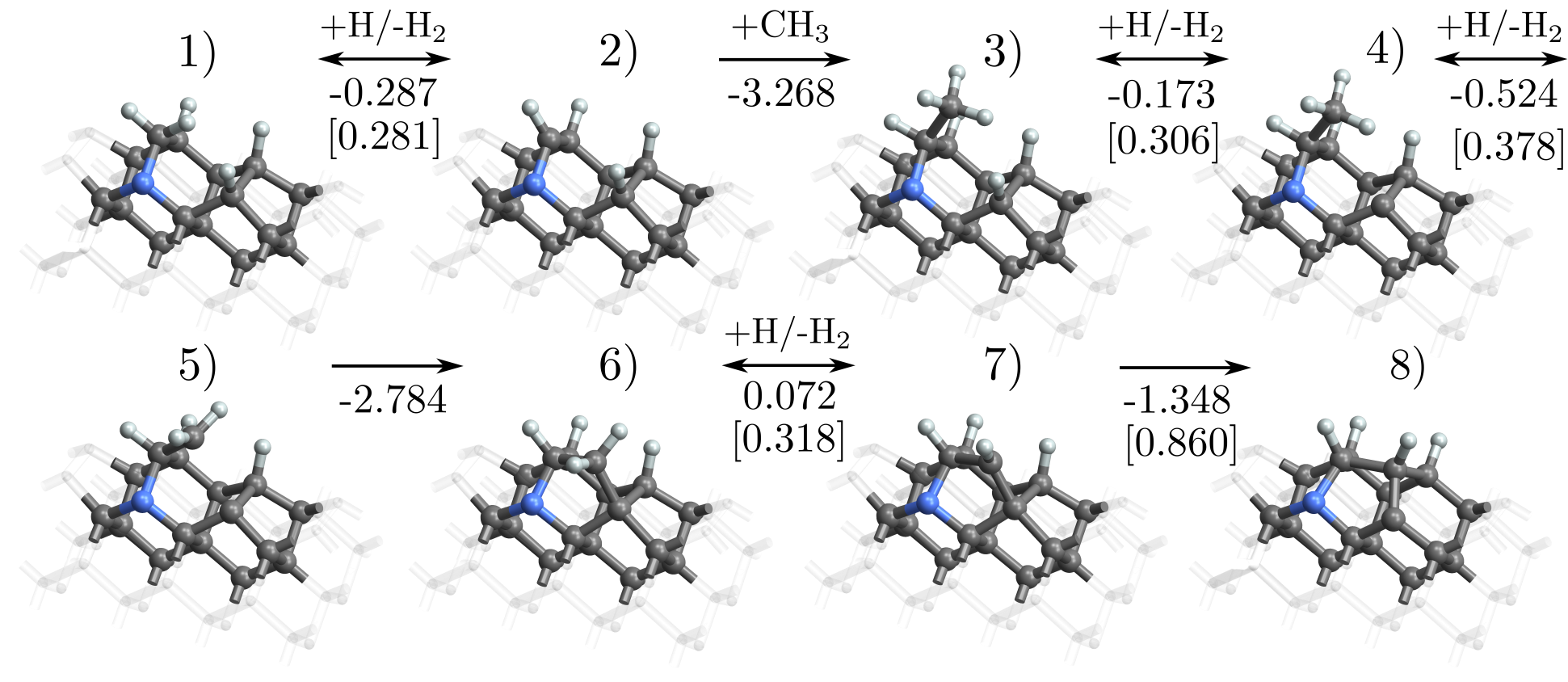}
	\caption{Mechanism for C insertion into a surface dimer adjacent to an incorporated C. This results in the formation of a new dimer unit on the surface and the beginning of a new layer. Reaction energies (in eV) are presented between each step in the reaction, and where applicable we also include barrier energies in square brackets.}
	\label{fig:dimer}
\end{figure*}

Figure~\ref{fig:dimer} presents the reaction mechanism and energy requirements for the insertion of C adjacent to the C already incorporated into the C--N dimer. The result is the formation of a new C--C dimer on the surface. This mechanism follows that identified by Cheesman~\textit{et al.} for formation of the dimer unit in the absence of N. Our calculations have similar energy requirements with the exception of a secondary dimer/opening closing mechanism as discussed below\cite{Cheesman2008a}.

In step $1\rightarrow2$, H abstraction forms a radical site on the previously incorporated C. This C adopts a bonding geometry similar to a free methyl radical, indicating that the N atom does not donate an electron and maintains its lone pair. This allows for the stable adsorption of CH${}_3$ in step $2\rightarrow3$ with no reaction barrier. Steps $3\rightarrow4$ and $4\rightarrow5$ consist of H abstractions on the adsorbed methyl and a H on the adjacent C--C dimer. N maintains its lone pair throughout these processes. This permits the CH${}_2$ adsorbate to bond at the adjacent surface site. Further H abstraction on the bridging CH${}_2$ produces a reactive radical. In step $7\rightarrow8$, the C--C dimer bond is severed to accommodate C insertion into the dimer. This is also a ring opening/closing mechanism and has an activation barrier commensurate with that observed for C insertion on a lone C--C dimer (0.860~eV vs 0.959~eV, c.f. Table~\ref{tab:results}).

In the work of Cheesman~\textit{et al.}, step $7\rightarrow8$ was considered in the absence of N and found to require a substantially lower activation barrier of 0.324~eV\cite{Cheesman2008a}. As this suggests that N hinders the formation of the dimer unit, we have also calculated step $7\rightarrow8$ in the absence of N using PBE functionals and our slab geometry. However, we instead find a transition state energy of 0.898~eV which indicates that N has little impact on ring opening/closing in step $7\rightarrow8$. As discussed previously, the deviation between our results and that presented in Cheesman~\textit{et al.} is likely attributable to over-relaxation of the transition state due to their cluster model.

The new C--C dimer is highly stable and readily adsorbs a further H radical to satiate the final dangling bond seen in state 8 of Figure~\ref{fig:dimer} ($\Delta E = -4.179$~eV). Note in particular that N does not partake in any immediate chemistry throughout dimer formation and maintains its lone pair. All reactions possess similar energetic requirements for abstraction and adsorption as that for a N-free surface (c.f. Table~\ref{tab:results}), and consequently N does not appear to benefit or hinder dimer formation.

\subsection{C addition between dimer rows}

Following formation of a new surface dimer, layer growth may propagate between dimer rows (often denoted the ``trough''). The addition of C across the trough is the beginning of a new surface dimer adjacent to the one previously formed, and therefore represents the first step in establishing a new dimer row. Once again, we identify that surface-embedded N does not hinder or benefit this process as it does not directly participate in bonding.

\begin{figure*}[]
	\centering
	\includegraphics[width=1\textwidth]{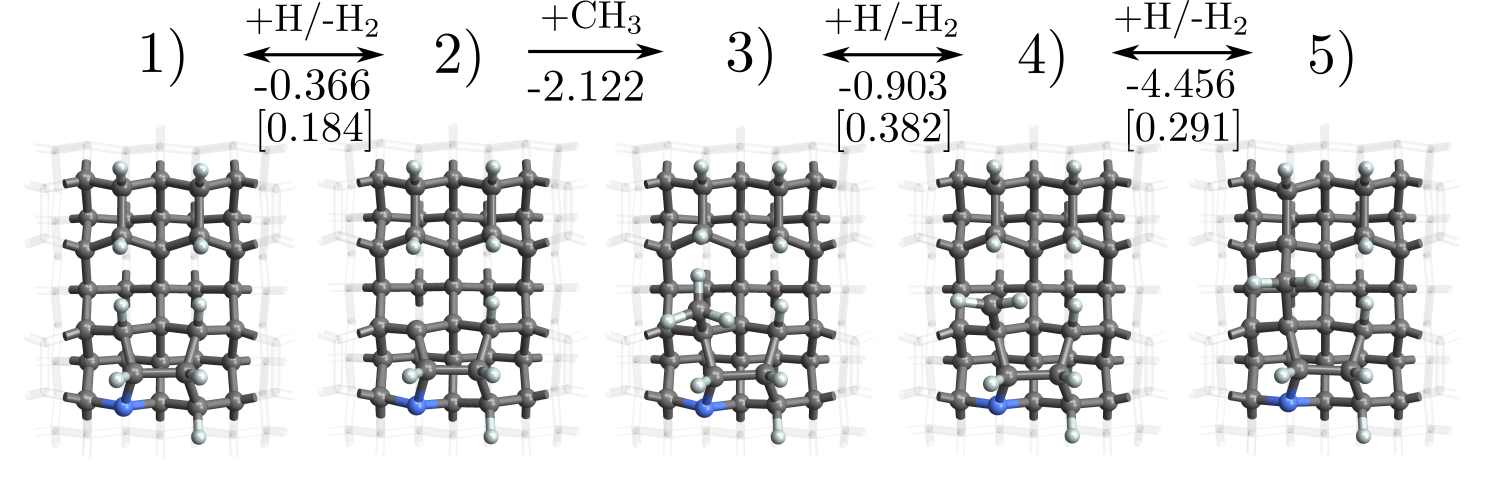}
	\caption{Mechanism for C addition between two dimer rows. Also known as the ``trough bridging'' mechanism, this begins the formation of a new dimer row along the C(100) surface. Reaction energies (in eV) are presented between each step in the reaction, and where applicable we also include barrier energies in square brackets.}
	\label{fig:trough}
\end{figure*}

The reaction mechanism for C addition across the trough including reaction and barrier energies is presented in Figure~\ref{fig:trough}. In step $1\rightarrow2$, H is abstracted from a ring-closed C on the surface layer. This permits the adsorption of CH${}_3$ onto the radical site in step $2\rightarrow3$. Two further H abstractions may then occur on the CH${}_3$ adsorbate (step $3\rightarrow4$) and on H from a C on the adjacent dimer row (step $4\rightarrow5$). The latter H abstraction results in the immediate bridging of the trough by the CH${}_2$ adsorbate. This bridging reaction also represents the first half of CH${}_2$ migration between dimer rows. Alternatively, through a mechanism similar to that presented in Figure~\ref{fig:dimer}, a new dimer can be formed across the trough. 

The addition of C across the trough appears to be a facile mechanism for propagating dimer rows, requiring only two successive H abstractions following CH${}_3$ adsorption. It is therefore a possible candidate for an atomic process which drives step-flow growth. Regardless, although the energies we determine in Figure~\ref{fig:trough} are similar to those identified by Cheesman~\textit{et al.} for the analogous process without N\cite{Cheesman2008a}, they argue that the trough bridging mechanism is not substantially faster than C insertion into a C--C dimer.

Through the three processes considered above, C insertion into a C--C dimer, formation of a new dimer, and C addition across dimer troughs, layer growth may propagate across the surface. When surface-embedded N is not directly involved in the formation of new bonds and growth, it does not appear to hinder or benefit growth processes. Instead, the N maintains its stable electron lone pair rather than donating charge to radical acceptor sites on the surface. The exception to this typical growth behaviour is C addition in the trough adjacent to N itself, which we treat in the next section.

Our results indicate that surface-embedded N catalyzes new-layer nucleation during step-flow modes but does not necessarily increase the rate of step-flow growth itself. Consequently, it may be expected that high N concentrations would result in a growth mode dominated by nucleation. Indeed, this effect is observed in experimental results in which increasing the N concentration leads to morphology changes indicating a transition from step-flow to nucleation dominated growth\cite{Achard2007}.

\section{Encapsulation of N}
\label{sec:III}

\begin{figure*}[]
  \centering
  \includegraphics[width=1\textwidth]{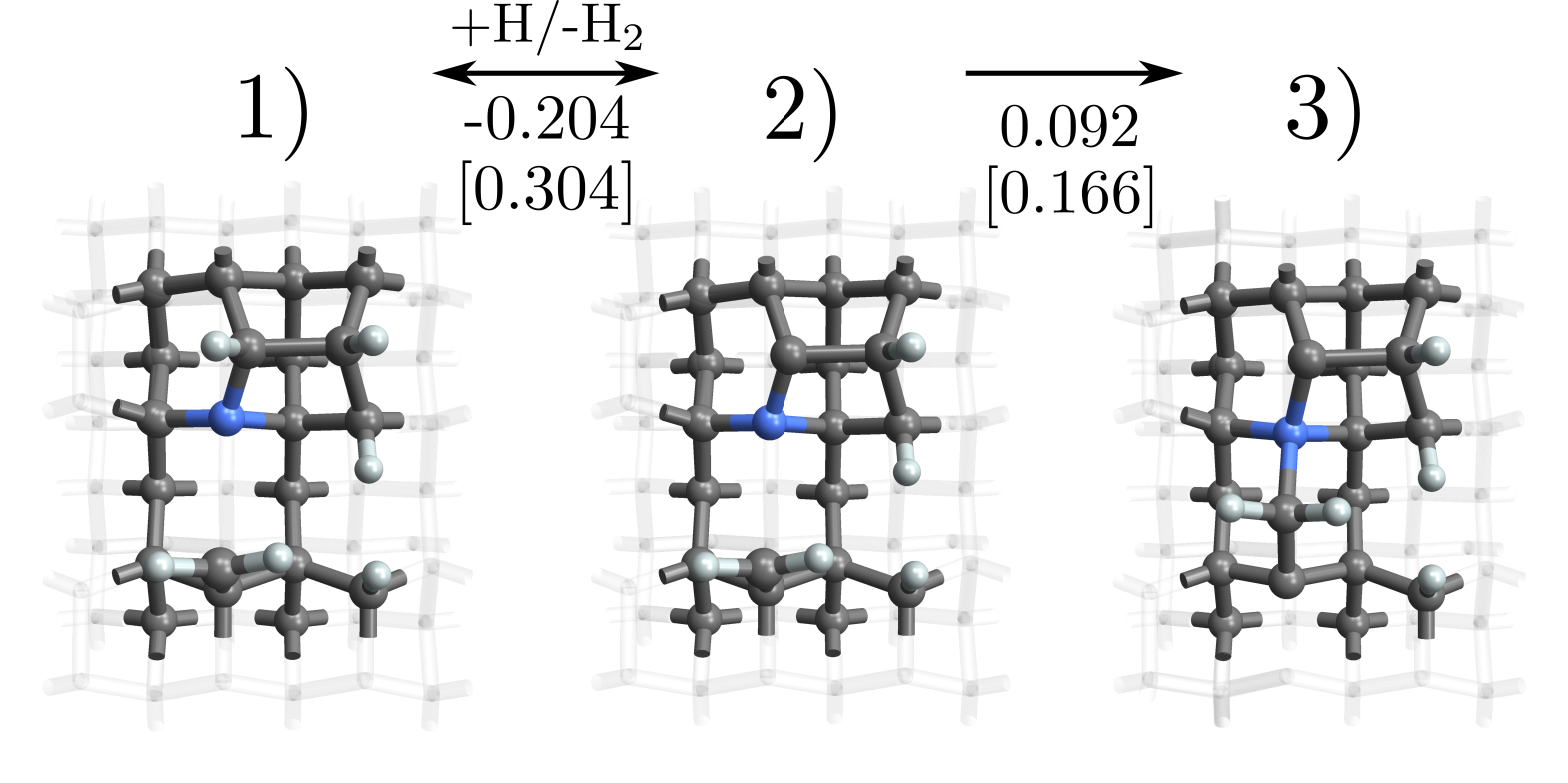}
  \caption{Encapsulation of surface-embedded N through C addition between two dimer rows. In state 1, the bridging CH${}_2$ radical is unable to bond with surface-embedded N. The N lone pair prevents formation of a C--N covalent bond. However, abstraction of a nearby surface H (step $1\rightarrow2$) creates an acceptor site permitting charge migration of a N electron. This breaks the N lone pair and allows for covalent bonding with the bridging CH${}_2$ in step $2\rightarrow3$. Reaction energies (in eV) are presented between each step in the reaction, and where applicable we also include barrier energies in square brackets.}
  \label{fig:penultimate}
\end{figure*}

Surface growth may therefore proceed unimpeded until the encapsulation of surface-embedded N by the new layer. A four-fold N bonding configuration requires C addition across the trough between the N and the C in the adjacent dimer row. However, a trough bridging process analogous to that depicted in Figure~\ref{fig:trough} is not viable. This scenario is depicted in Figure~\ref{fig:penultimate}. In state 1, a radical CH${}_2$ adsorbate attempts to form a covalent bond with surface-embedded N. However, our calculations indicate that such a bond is not energetically stable. Similarly, it is also not possible for a CH${}_3$ radical to adsorb directly onto surface-embedded N. Any covalent bonding is prohibited by the stability of the N lone pair.

We find that a stable bond between N and the bridging CH${}_2$ adsorbate is only possible following abstraction of a nearby surface H atom. For example, H abstraction on the newly formed dimer unit in step $1\rightarrow2$ of Figure~\ref{fig:penultimate}. This creates an acceptor site for a migrating charge from the N lone pair. In step $2\rightarrow3$, the CH${}_2$ radical is able to bridge the trough and form a stable covalent bond with surface-embedded N. Encapsulation of surface-embedded N defect therefore requires breaking its lone pair and charge migration to a surface acceptor site. While state 3 is stable, reaction $2\rightarrow3$ is endothermic and indicates that the CH${}_2$ adsorbate preferentially remains radical. This reflects the stability of the N lone pair.

\begin{figure*}[]
	\centering
	\includegraphics[width=1\textwidth]{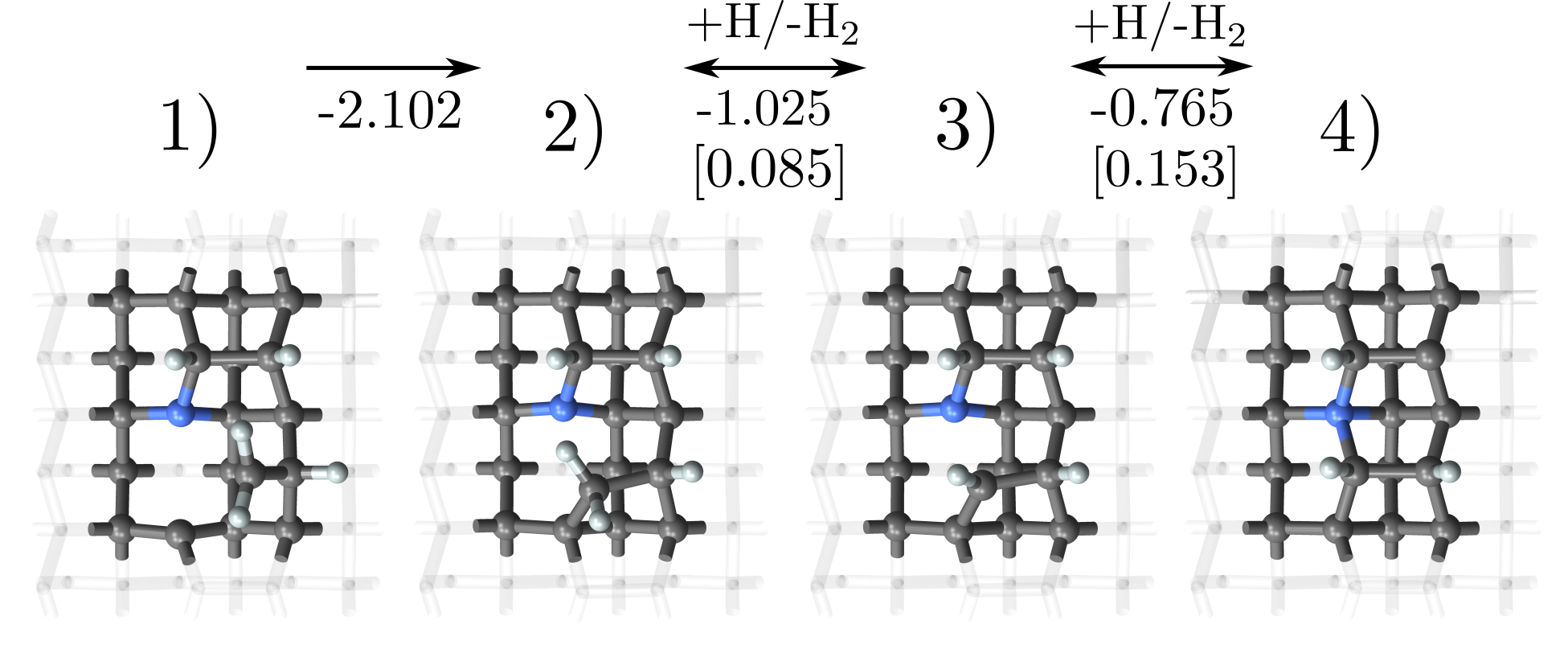}
	\caption{Encapsulation of surface-embedded N. Here we consider complete growth of a new surface layer with the exception of a final bond between the surface-embedded N and a bridging C. Bonding only occurs if the N lone pair is broken through donation of an electron to an acceptor site (step $3\rightarrow4$).}
	\label{fig:final}
\end{figure*}

Exothermic encapsulation of surface-embedded N can still occur through a different reaction pathway. Consider state 1 depicted in Figure~\ref{fig:final}. In this scenario, growth of a new layer is complete with the exception of encapsulation of surface-embedded N. Methyl adsorption and subsequent H abstractions result in a CH${}_2$ adsorbate attempting to bridge the final trough between a C and the surface-embedded N. In step $1\rightarrow2$, the bridging C adsorbs to the sub-surface C (analogous to step $5\rightarrow6$ in Figure~\ref{fig:dimer}). In step $2\rightarrow3$, H abstraction on the bridging C creates a radical site ready to bond with the surface-embedded N. During this mechanism the remaining H on the bridging C rotates away from the N lone pair, greatly reducing steric interactions. Consequently, step $2\rightarrow3$ has a low barrier energy of 0.085~eV and results in a 1.025~eV increase in stability.

Once again, encapsulation of surface-embedded N requires breaking the lone pair to accommodate covalent bonding with the C radical. This is only possible if a lone-pair electron can be donated to a suitable acceptor site. In step $3\rightarrow4$ this is realised through H abstraction of a surface H. In this circumstance N donates an electron to the acceptor site and bonds with the bridging C immediately following abstraction. This fully encapsulates the surface-embedded N and forms a positively-charged defect and negatively-charged acceptor site on the surface. Alternatively, given that during CVD growth anywhere between $1-10$\% of the surface consists of acceptor sites rather than H termination (depending on growth conditions\cite{Butler2009}), step $3\rightarrow4$ may occur immediately without further H abstraction. In contrast to step $2\rightarrow3$ in Figure~\ref{fig:penultimate}, N encapsulation is exothermic by 765~meV. This can be attributed to the strain of the C--C dimer bond of state 3 in Figure~\ref{fig:final}. Not only does N encapsulation satisfy all radicals, it results in the formation of an unstrained surface dimer.

The facile pathway from surface-embedded to substitutional N identified in Figure~\ref{fig:final} provides a potential explanation for the low yield of NV centers during CVD growth. Experimentally, it has been demonstrated that at most 0.5\% of incorporated N forms NV centers\cite{Lobaev2018}. Experimental results suggest that NV centers form as a unit on (110) surfaces, with N incorporating into the lattice first before the the vacancy is produced in the overgrown layer\cite{PhysRevB.86.035201}. \textit{Ab initio} calculations support this hypothesis and predict a similar mechanism for NV formation on the (100) surface\cite{PhysRevB.88.245301}.

This would therefore suggest that the overgrowth mechanism in Figure~\ref{fig:final} is a critical point which determines either formation of substitutional N or an NV center. If the CH${}_2$ adsorbate were to be prevented from bridging the final trough (or CH${}_3$ initially adsorbing to the surface), it is conceivable that a vacancy would form during overgrowth of the next layer (the second layer). While we do not identify any such vacancy formation mechanism in this work, one possibility could be if growth of the second layer occurs simultaneously with the first layer. The step-edge of the secondary layer could then overgrow the local surface depicted in Figure~\ref{fig:final} before the reaction can proceed, subsequently forming an NV center. However, given the low energy requirements for N encapsulation in Figure~\ref{fig:final} this is unlikely, thus giving rise to the low NV to substitutional N ratios observed experimentally.

\section{Conclusion}

Three decades of \textit{ab initio} research has produced no consensus on the atomistic mechanism for N catalysis of CVD diamond growth. In this work we identify a new catalytic effect relevant to step-flow dominated growth modes on (100) surfaces. Specifically, surface-embedded N drastically enhances the rate of new layer nucleation by reducing the energy barrier for C insertion. The presence of N increases the rate of the ring opening/closing reaction by a factor of 400, the greatest enhancement to any key (100) diamond growth process to date. Experimental support for our nucleation model may be obtained through correlation of surface-embedded N and nucleated layers following CVD. One way to achieve this could be use of an NV-based quantum microscope to locate N defects in the surface through their hyperfine structure\cite{Thiel973}. Conventional surface imaging techniques such as STM could then be used to identify surrounding layer growth which has propagated from the defect\cite{Bobrov2001}.

This work has also established the first atomic model describing overgrowth of N during (100) diamond CVD. We determine that surface-embedded N does not impede typical layer growth processes. N maintains its strong electron lone pair and does not participate in bonding during formation of the dimer unit or C bridging of non-adjacent troughs. The exception is encapsulation of surface-embedded N to form a sub-surface defect through C bridging on the adjacent trough. This reaction requires donation of a lone pair electron to a surface acceptor site. Moreover, the process may be endothermic or exothermic depending on the presence of surrounding new-layer growth. The resulting sub-surface defect is four-fold coordinated and positively charged.

\section*{Acknowledgments}
We acknowledge funding from the Australian Research Council (DE170100169). This research was undertaken with the assistance of resources and services from the National Computational Infrastructure (NCI), which is supported by the Australian Government.

\bibliography{paper}

\end{document}


\title{\vspace{-2cm}Supplementary material: An atomistic model for nitrogen catalysis during diamond CVD growth}
 \author{ Lachlan M. Oberg$^1$, Marietta Batzer$^2$, Alastair Stacey${}^3$, and Marcus W. Doherty${}^{1,4}$}
\date{%
    $^1$Laser Physics Center, Research School of Physics, Australian National University, Australian Capital Territory 2601, Australia\\%
    $^2$Department of Physics, University of Basel, Klingelbergstrasse 82, Switzerland \\
    $^3$School of Science, RMIT University, Melbourne, Victoria 3001, Australia \\
    $^4$Quantum Brilliance Pty Ltd, 116 Daley Road, Acton, Australian Capital Territory 2601, Australia\\[2ex]%
    \today
}
\maketitle

\section*{Introduction}

In this supplementary material we demonstrate the accuracy of our \textit{ab initio} simulations and validate our slab model of the H-C(100) diamond surface. All calculations are performed using version 5.4.4 of the Vienna \textit{Ab initio} Simulation Package (VASP) using the projector augmented wave method\cite{Kresse1996a,Kresse1994a,Kresse1996,Joubert1999} with PBE functionals\cite{Perdew1996b}. In Section~\ref{sec:bulk} we determine the plane wave cut-off energy required to adequately simulate bulk diamond and optimize the lattice constant for all further calculations. In Section~\ref{sec:vertical} we investigate the relationship between slab thickness, the H-C(100) work function, and the energetics of C insertion into the C--N dimer. In Section~\ref{sec:lateral} we investigate the relationship between the lateral dimensions of the slab, intra-cellular interactions between N defects, and the energetics of C insertion into the C--N dimer. Using these results we identify the slab dimensions required to adequately represent the physics and chemistry of the diamond surface. Finally, in Section~\ref{sec:instab} we further investigate the instability of state 5 during C insertion into the C--C dimer.

\section{Bulk diamond}
\label{sec:bulk}

\subsection{Energy cut-off}

VASP was used to determine the optimal energy cut-off for bulk diamond. The energy cut-off determines the size of the plane-wave basis and the minimum value possible must be chosen in order to maximise computational efficiency. To determine this cut-off, the diamond lattice constant was fixed to its experimental value of 3.567 \AA\cite{Madelung2004} and the $\mathbf{k}$-point sampling was chosen to be a $8\times8\times8$ Monkhorst-Pack mesh. The tolerance for electronic convergence was set to 0.1~meV.

\begin{figure}[]
	\centering
	\includegraphics[width=0.6\textwidth]{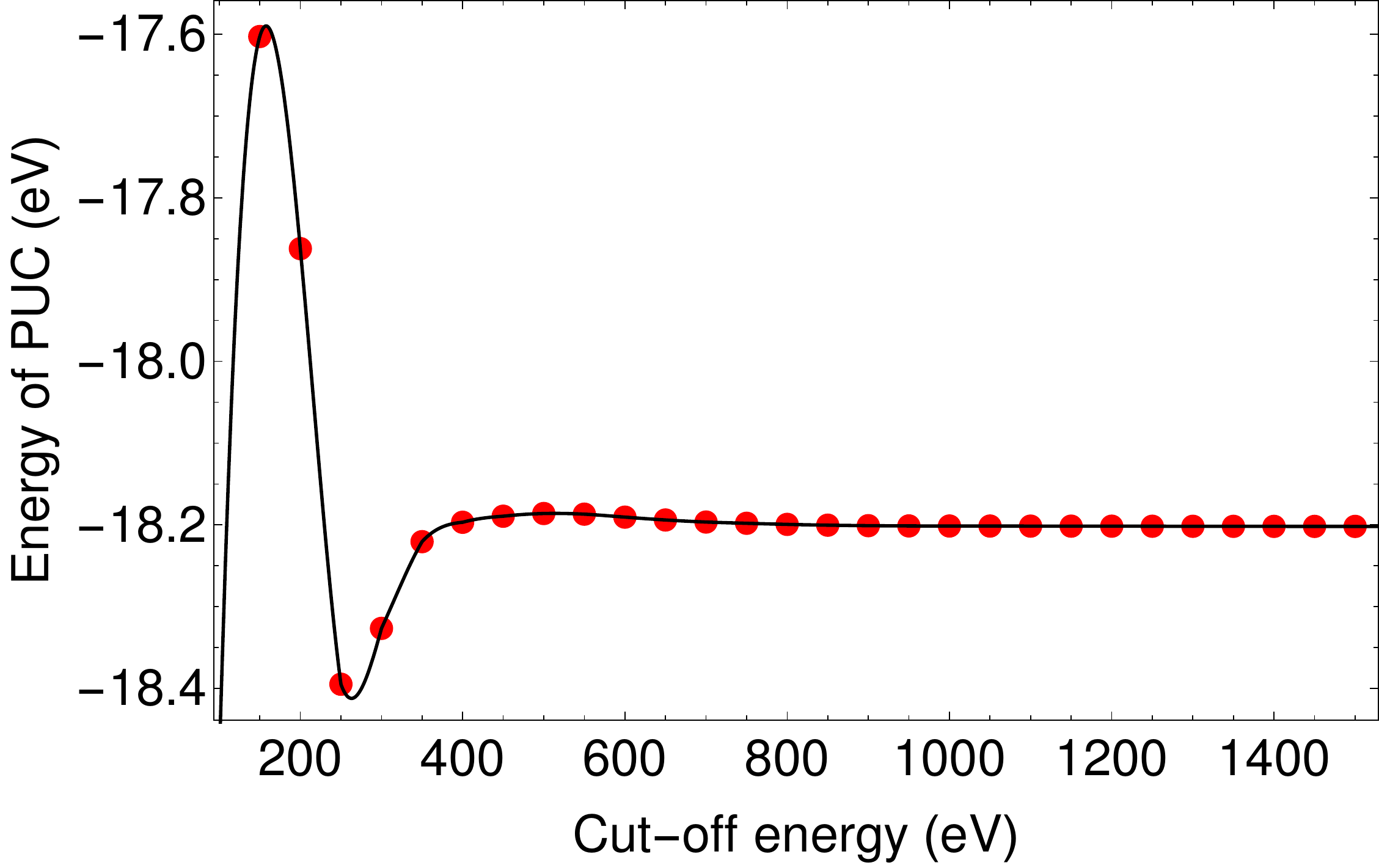}
	\caption{Cut-off energy vs energy of a primitive two-atom diamond unit cell. The energies were determined using VASP, assuming the experimental lattice constant of 3.567 \AA, a $8\times8\times8$ Monkhorst-Pack $\mathbf{k}$-point mesh, and a convergence tolerance of 0.1 meV.}
	\label{fig:cut-off}
\end{figure}

The energy of the two-atom bulk primitive unit cell (PUC) was calculated as a function of energy cut-off, and the relationship is displayed in Figure~\ref{fig:cut-off}. As can be seen, the energy of the unit cell rises between energy cut-offs of 250 eV and 500 eV. Further increasing the energy cut-off results in a slight decrease of the unit-cell energy, before converging to a constant value of $\sim18.2$~eV beyond 800 eV. However, a cut-off energy of 600 eV yields a unit cell energy only 10 meV above the converged value (a mere relative difference of 0.06\%). Hence, given that 99.94\% of the converged energy can be reproduced with a large 25\% reduction of the basis size, we choose a 600~eV cut-off energy for all future calculations.

\subsection{Lattice constant}

\begin{figure}[]
	\centering
	\includegraphics[width=0.6\textwidth]{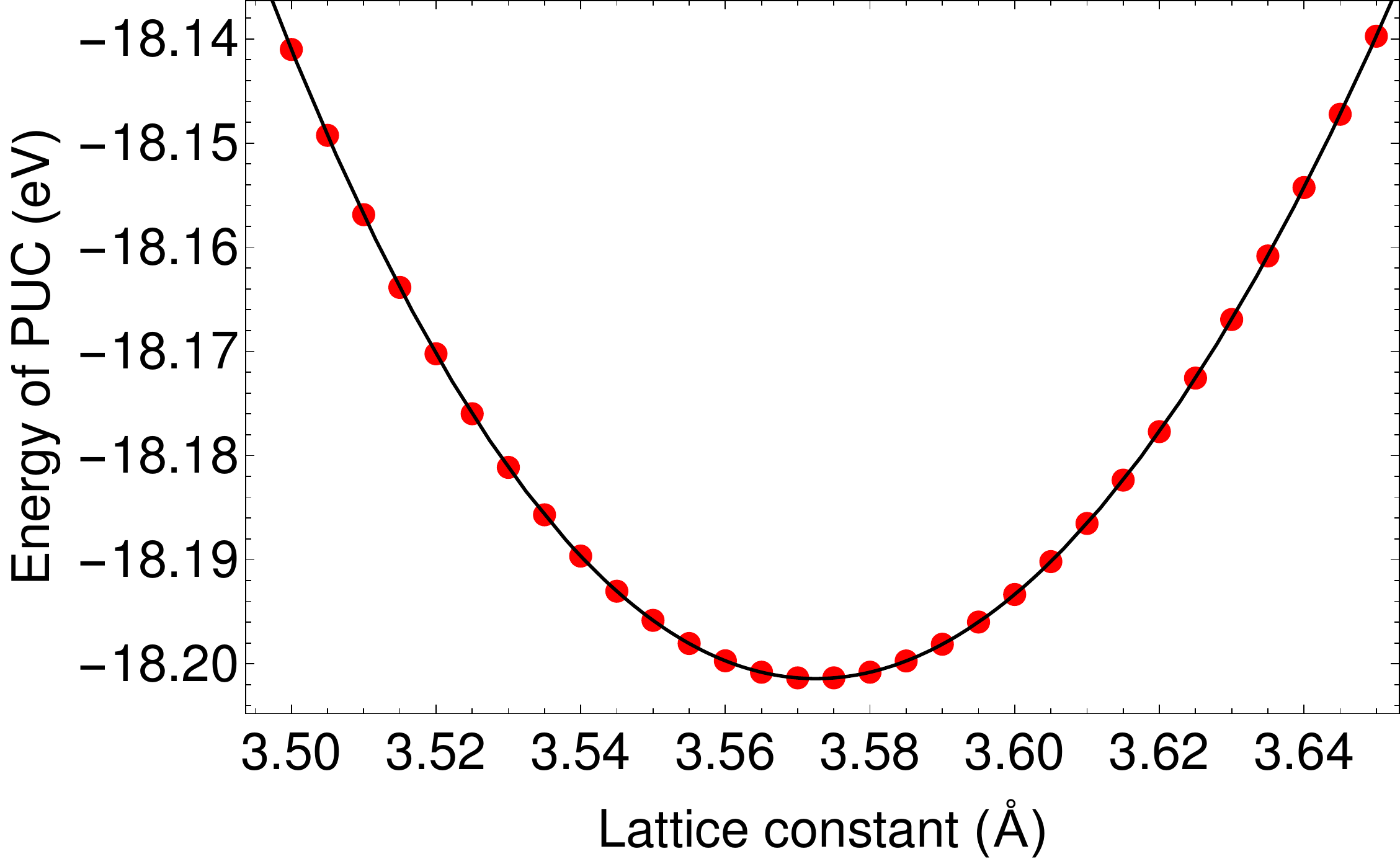}
	\caption{Lattice constant vs energy of a primitive two-atom diamond unit cell. The energies were determined using VASP, with an energy cut-off of 600~eV, an $8\times8\times8$ Monkhorst-Pack sampling and a convergence tolerance of 0.1 meV. The results indicate a equilibrium lattice constant of 3.573~\AA, within 0.25\% of the experimental value\cite{Madelung2004}.}
	\label{fig:latticeConst}
\end{figure}

The lattice constant resulting in the minimum energy of the bulk diamond PUC was determined using VASP. We use an energy cut-off of 600~eV and $\mathbf{k}$-point sampling of $8\times8\times8$ in the Monkhorst-Pack scheme. The results are displayed in Figure~\ref{fig:latticeConst}, and indicate an equilibrium lattice constant of $3.573$ \AA, within 0.25\% of the experimental value\cite{Madelung2004}. This value is used for all future calculations.

\section{Optimisation of slab thickness}
\label{sec:vertical}

The bulk PUC was used to construct slabs of the reconstructed (2x1) H-C(100) surface of varying dimensions\cite{Furthmuller1996}. The computational methodology used is identical to that presented in the main text.

\subsection{Work function}
All surfaces possess a work function, which specifies the energy of the Fermi-level ($E_f$) with respect to the vacuum energy ($E_\text{vac}$) as per
\begin{align}\label{WF}
W = E_\text{vac} - E_f.
\end{align}
For semiconductors, this represents the energy required to relocate an electron from the valence band maximum (VBM) to the vacuum\cite{C.2004}. The work function therefore provides a reference frame for comparing binding energies and therefore the strength of chemical bonds. The value of $W$ as determined by \textit{ab initio} calculations is a function of slab thickness. Hence, the thickness of the slab must be chosen to provide a good representation of the surface's electronic properties. 

\begin{figure}[]
	\centering
	\includegraphics[width=0.6\textwidth]{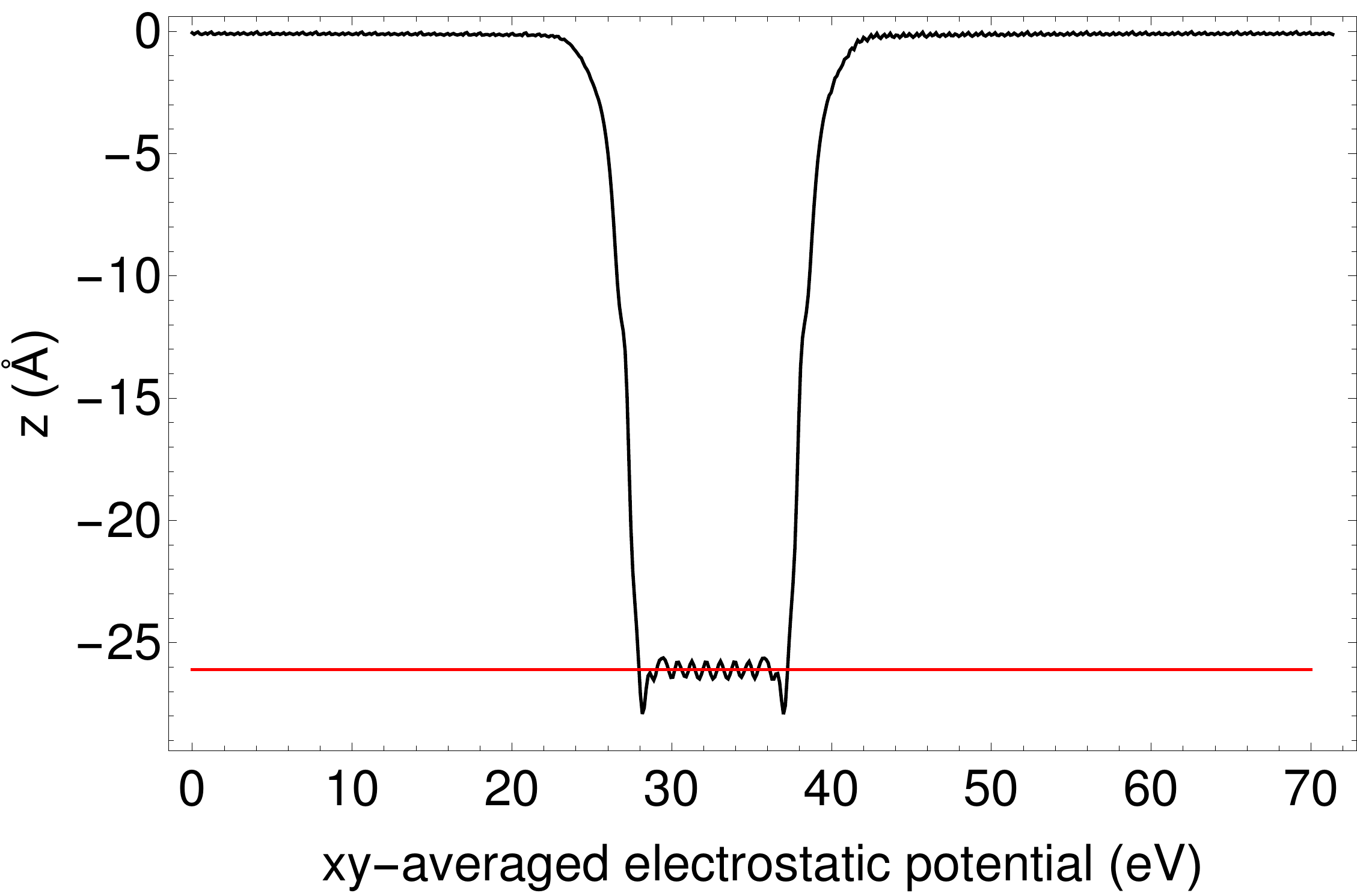}
	\caption{Averaged $xy$-planar electrostatic potential vs $z$-coordinate in a four PUC thick H-C(100) slab supercell. All energies have been scaled such that the vacuum level resides at 0 eV. The slab is positioned between 30 and 40~\AA \ into the supercell, indicated by the presence of eight small oscillations in the potential (the carbon layers) bordered by larger peaks (the H terminations on either side of the surface). In red, the average potential of the carbon layers has been plotted at $-26.2$~eV.}
	\label{fig:aveElec9LayerH}
\end{figure}

To obtain the vacuum level, VASP was used to produce the $xy$-planar averaged, local electrostatic potential of each slab. The resulting potential is therefore a function of the out-of-plane coordinate (i.e., the \textless100\textgreater \ direction), here defined as $z$. An example of this potential is shown in Figure~\ref{fig:aveElec9LayerH} for a H-C(100) slab which is four PUCs thick. All energies have been scaled such that the vacuum energy resides at 0~eV. The slab is positioned between 30 and 40~\AA and one can observe eight small oscillations in the potential (produced by the eight carbon layers) bordered by larger peaks (produced by the H-terminations on either surface). The average potential of the carbon layers sits deep below the vacuum level, indicated by the dotted red line at $-26.2$~eV.

The Fermi level can be identified as the valence band maximum (VBM) of the slab band structure. While the VBM has a well-defined position with respect to the vacuum level, this position is distorted by surface induced effects. For example, the surface termination introduces additional states into the valence band structure, while the slab geometry itself causes quantum confinement effects which modulate band energies\cite{Stoneham1975}. Thus, the VBM is not consistent for each slab, and so cannot be used to meaningfully compare work functions between different slab thicknesses. Instead, an independent reference energy for the Fermi level is required.

This reference energy can be obtained from bulk diamond, which does not suffer from surface induced effects. In particular, consider the average electrostatic energy of the bulk-like carbon layers in the slab. As an example, this is indicated by the red line in Figure~\ref{fig:aveElec9LayerH} for a four PUC thick H-C(100) surface. This energy can be aligned with the average electrostatic energy of the carbon atoms in bulk diamond obtained via our simulations in Section~\ref{sec:bulk}. Following alignment, the vacuum level of the slab obtains a well-defined position with respect to the bulk Fermi level. The work function can then be calculated as the difference between these two values.

\begin{figure}[]
	\centering
	\includegraphics[width=0.6\textwidth]{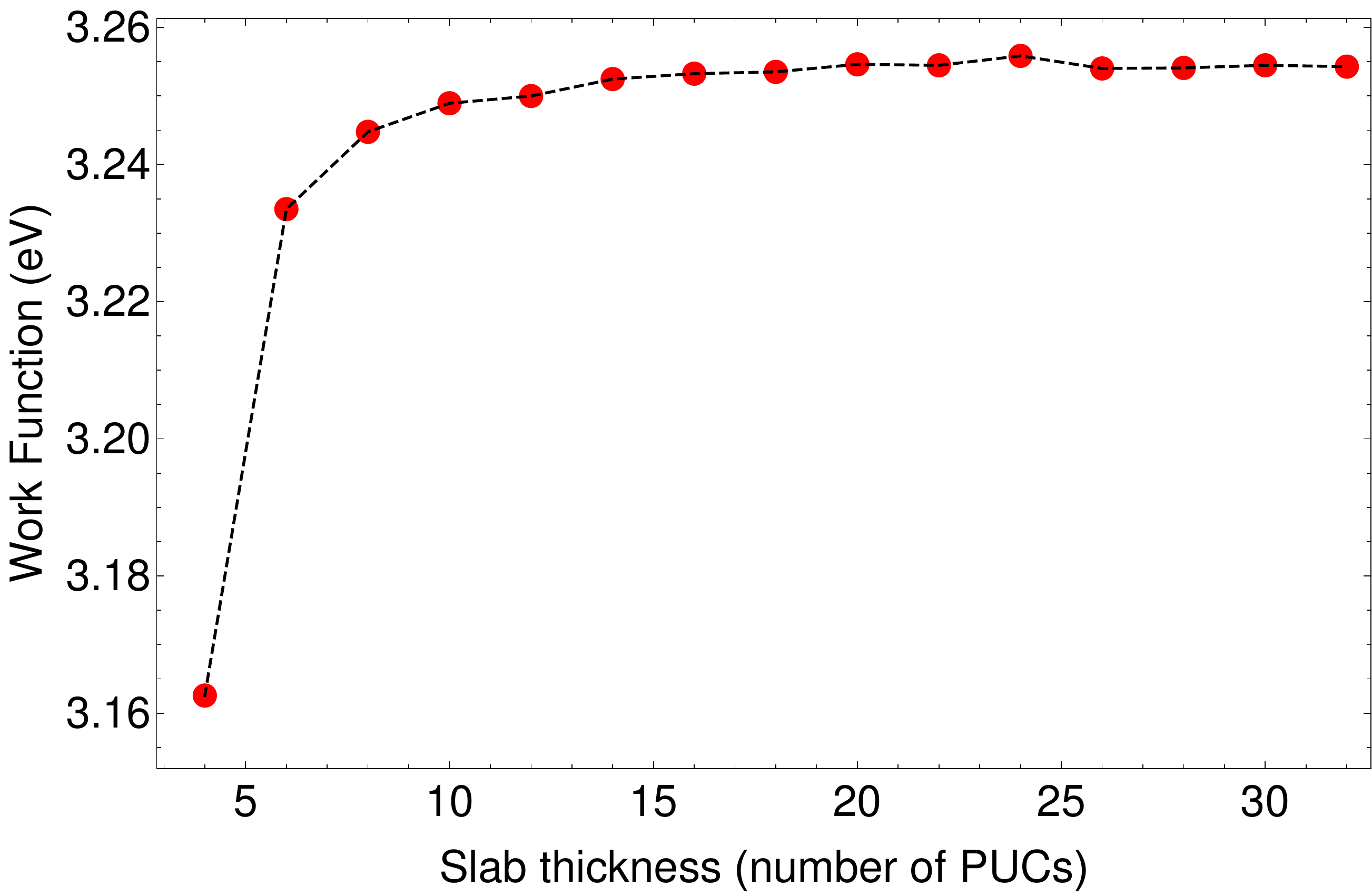}
	\caption[]{Work function vs thickness of slab for the H-terminated C(100) surface.}
	\label{fig:WFcalcs}
\end{figure}

This methodology is well established in solid-state DFT literature\cite{Walle1987,Alkauskas2008} and was applied for our slab calculations. In Figure~\ref{fig:WFcalcs}, we depict the relationship between slab thickness and work function for a H-terminated C(100) slab. The results indicate that the work function converges to a constant value of $\approx3.25$~eV for slabs 10 PUCs or greater thick. However, even a thickness of six PUCs produces a work function within 20~meV of this value. 

PBE functionals are sufficient to determine the point at which work function convergence occurs. However, due to the band gap problem the converged energies will severely underestimate the experimental values. This can be resolved through the use of hybrid functionals or GW corrections. Due to the computational demand of the former, the position of the slab VBM and conduction band minimum (CBM) can instead be rescaled according to accepted GW values through the Delta SCF method. In particular, a recent review by Chen and Pasquarello found that the GW VBM and CBM energies are shifted by $-0.66$~eV and $+0.88$~eV with respect to the PBE energies\cite{Chen2012}. Consulting Figure~\ref{fig:WFcalcs}, our calculations therefore yield a converged work function of 3.91~eV for H-terminated diamond.

\subsection{C insertion into C--N}
\begin{figure}
    \centering
    \begin{subfigure}[b]{0.4\textwidth}
        \includegraphics[width=\textwidth]{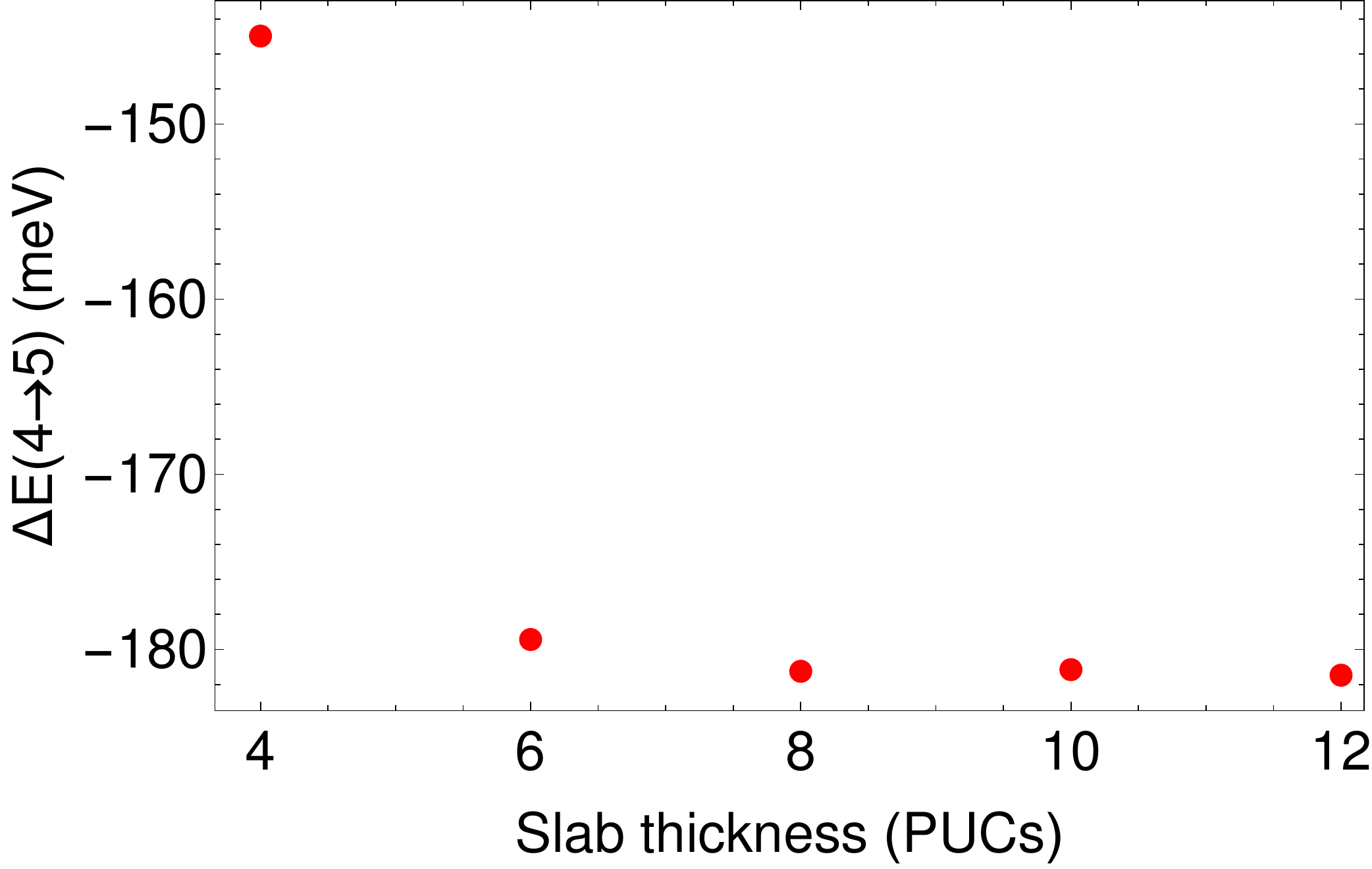}
        \caption{$\Delta E(4\rightarrow5)$.}
    \end{subfigure}
    ~ 
    \begin{subfigure}[b]{0.4\textwidth}
        \includegraphics[width=\textwidth]{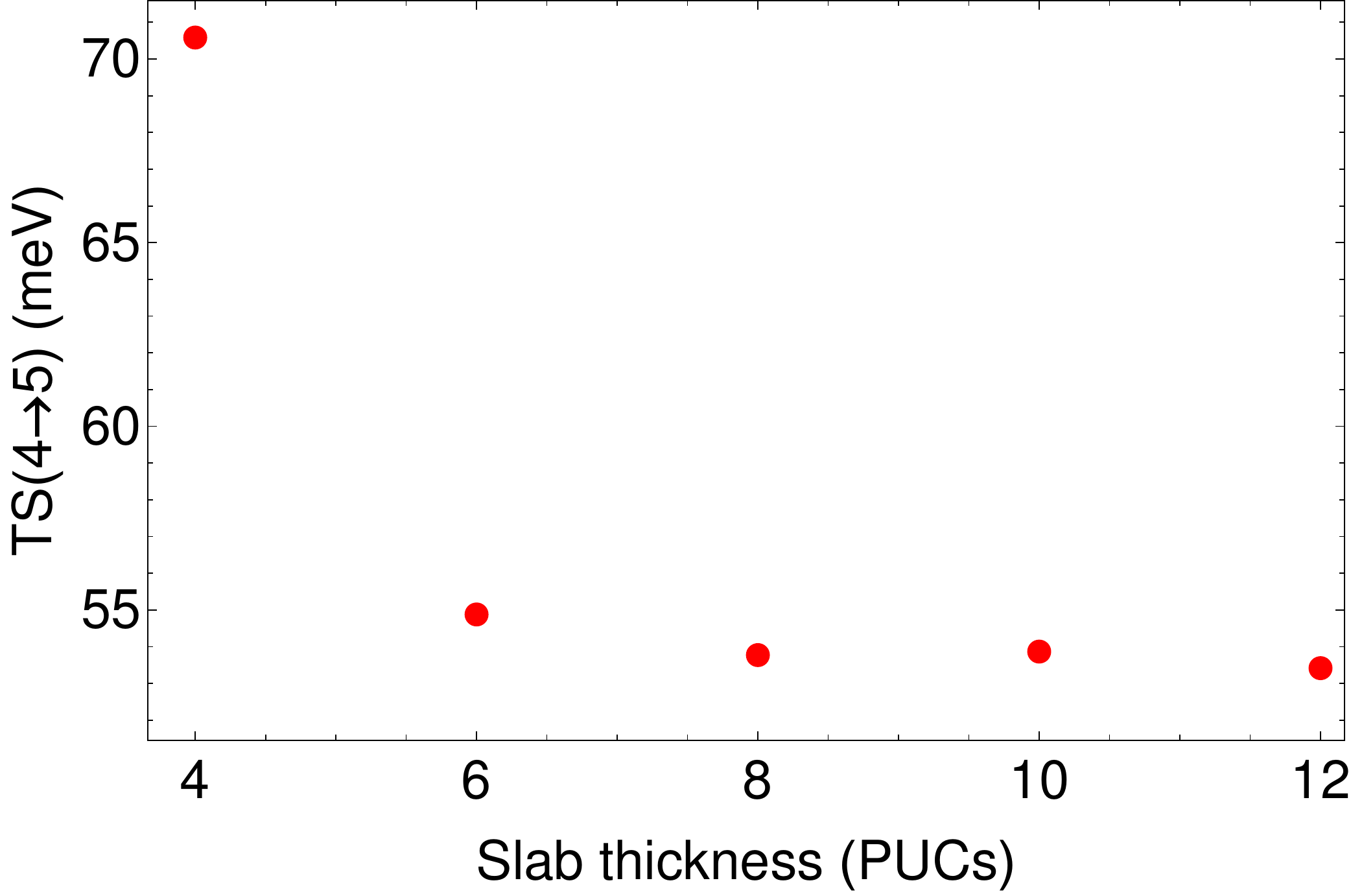}
        \caption{TS$(4\rightarrow5)$.}
    \end{subfigure}
    
    \begin{subfigure}[b]{0.4\textwidth}
        \includegraphics[width=\textwidth]{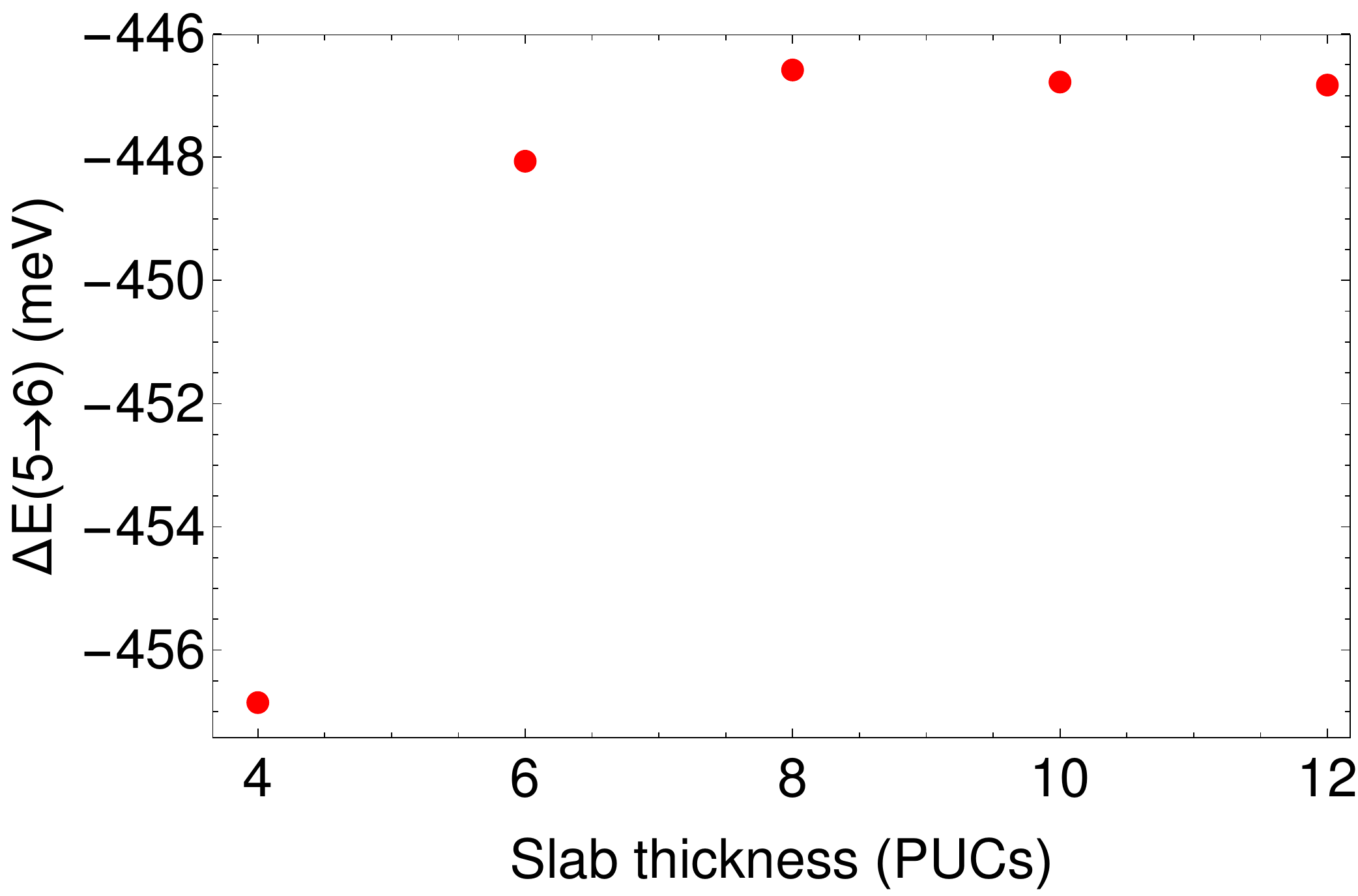}
        \caption{$\Delta E(5\rightarrow6)$.}
    \end{subfigure}
    ~
        \begin{subfigure}[b]{0.4\textwidth}
        \includegraphics[width=\textwidth]{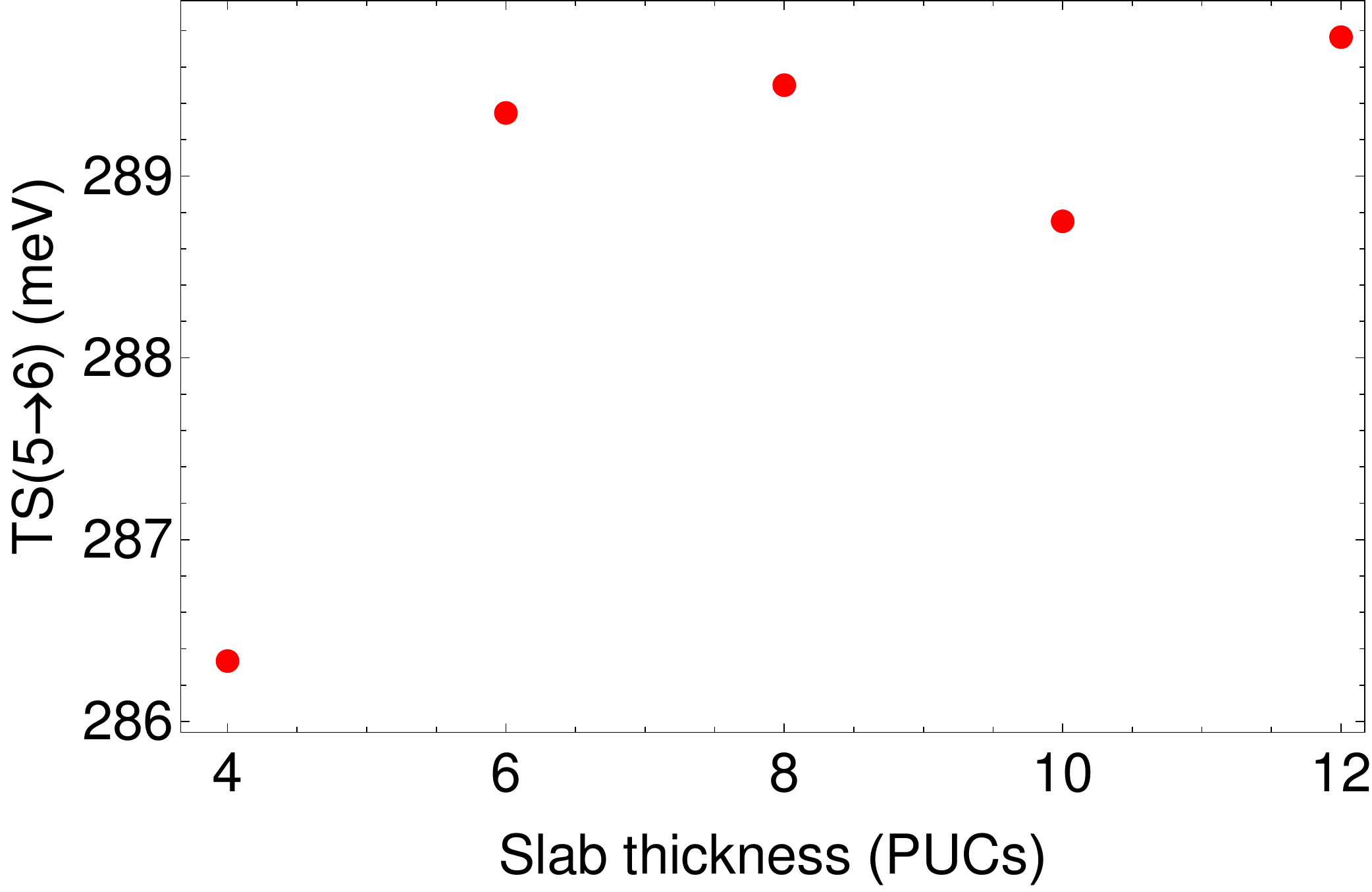}
        \caption{TS$(5\rightarrow6)$.}
    \end{subfigure}
    \caption{Energetics for steps $4\rightarrow5$ and $5\rightarrow6$ of C insertion into a C--N dimer as a function of slab thickness in PUCs. The lateral dimensions of the slab are $5\times4$ PUCs, as described in Section~\ref{sec:lateral}.}\label{fig:vert}
\end{figure}

We also sought to determine whether the trend observed in the work function is reflected in the energetics for diamond surface reactions. Consequently, we calculated both $\Delta E$ and the barrier energy (TS) for steps $4\rightarrow5$ and $5\rightarrow6$ of C insertion into a C--N dimer (as presented in the main text) as a function of slab thickness. Our results are presented in Figure~\ref{fig:vert} and indicate that convergence of surface energetics scales similarly to the work function. All energies differ by no more than several meV for slabs with a thickness between six and eight PUCs. At worst, the difference in energies are a mere 30 meV between the thinnest and thickest slabs considered. Given that six PUCs result in a work function within 20~meV of the converged value for the work function, and within 2~meV for the converged reaction and barrier energies for the ring opening/closing mechanism, this slab thickness was chosen for all future calculations. 

\section{Optimization of lateral slab dimensions}
\label{sec:lateral}

\begin{figure}[]
	\centering
	\includegraphics[width=0.6\textwidth]{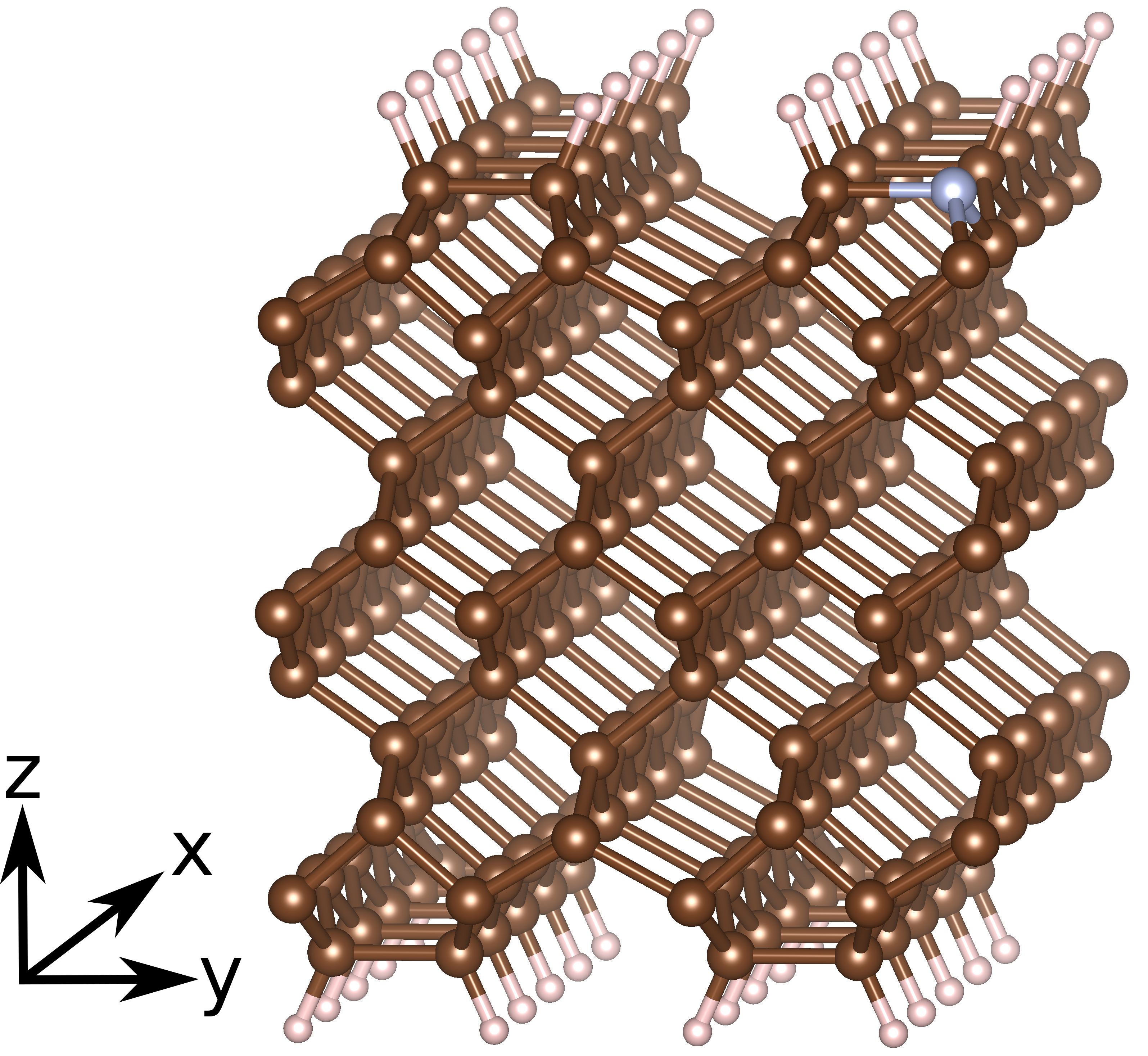}
	\caption{The optimized slab supercell of the H-C(100) surface, consisting of 240 C ions (brown) with dimensions $5\times4\times6$ in terms of bulk diamond PUCs. The surface is terminated on both sides by H ions (pink) with a single N defect embedded into the surface (blue).}
	\label{fig:surf}
\end{figure}

Ions in periodic slab models interact with ions in adjacent supercells to effectively reproduce geometric properties of the surface. However, this can also result in undesired intra-cellular interactions between single defects or chemical adsorbates which are intended to be studied in isolation. Consequently, we now seek to mitigate these interactions by studying the relationship between lateral slab dimensions and the resulting interaction energies. This shall be done in two ways. Firstly, we shall use the slab band structure to directly calculate interaction energies between surface-embedded N defects. Secondly, we will calculate the reaction and barrier energies for steps $4\rightarrow5$ and $5\rightarrow6$ of C insertion into the C--N dimer as a function of lateral slab dimensions. For the following discussion we assign the direction of the surface dimer bonds as the $y$ direction, while adjacent dimers form rows along the $x$ direction as depicted in Figure~\ref{fig:surf}.

\subsection{Intra-cell interaction energies}

In these simulations we consider a single N atom embedded in the surface of a H-C(100) slab (e.g., Figure~\ref{fig:surf}). The interaction energy between N defects in adjacent supercells was determined by calculating the slab band structure. We sample $\mathbf{k}$-points along the principal axes corresponding to the $x$ and $y$ directions in reciprocal space (i.e., between the $\Gamma$ and $X$ high symmetries in either lateral direction). The resulting wavefunctions are projected onto a Wigner-Seitz sphere centered on the N defect to determine the corresponding local density of states. This allows us to identify distinct energy bands which can be attributed to the N defect. As predicted by tight-binding theory\cite{Atkins2011b}, we find that the dispersion of these bands follows the relationship
\begin{equation}\label{band}
E(k) = \alpha + 2\beta\left( \frac{2\pi k}{\gamma} \right ),
\end{equation}
for some energy off-set $\alpha$, interaction energy $\beta$, and band width $\gamma$.

\begin{figure}
    \centering
    \begin{subfigure}[b]{0.45\textwidth}
        \includegraphics[width=\textwidth]{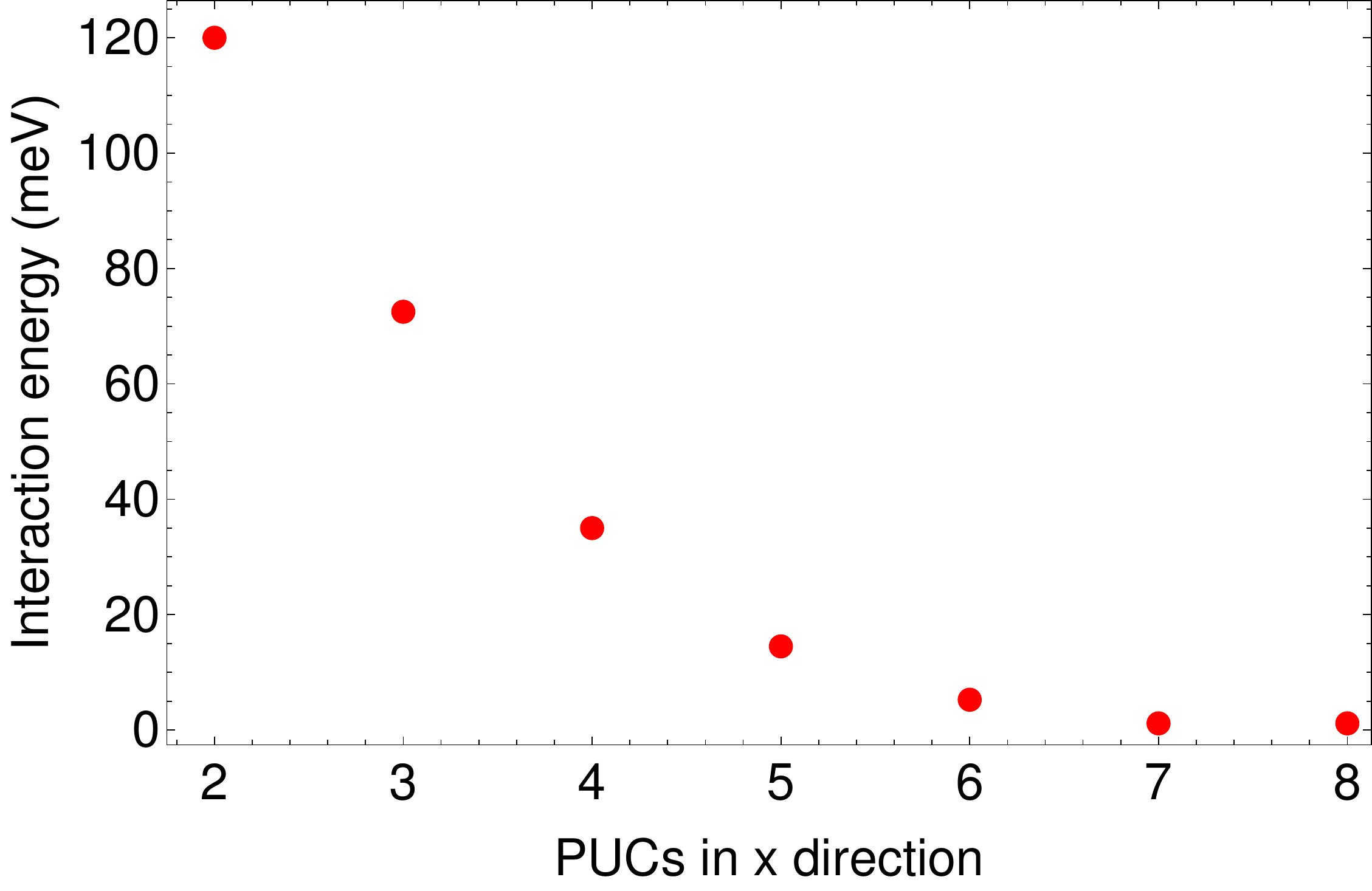}
        \caption{$x$ direction.}
    \end{subfigure}
    ~ 
    \begin{subfigure}[b]{0.45\textwidth}
        \includegraphics[width=\textwidth]{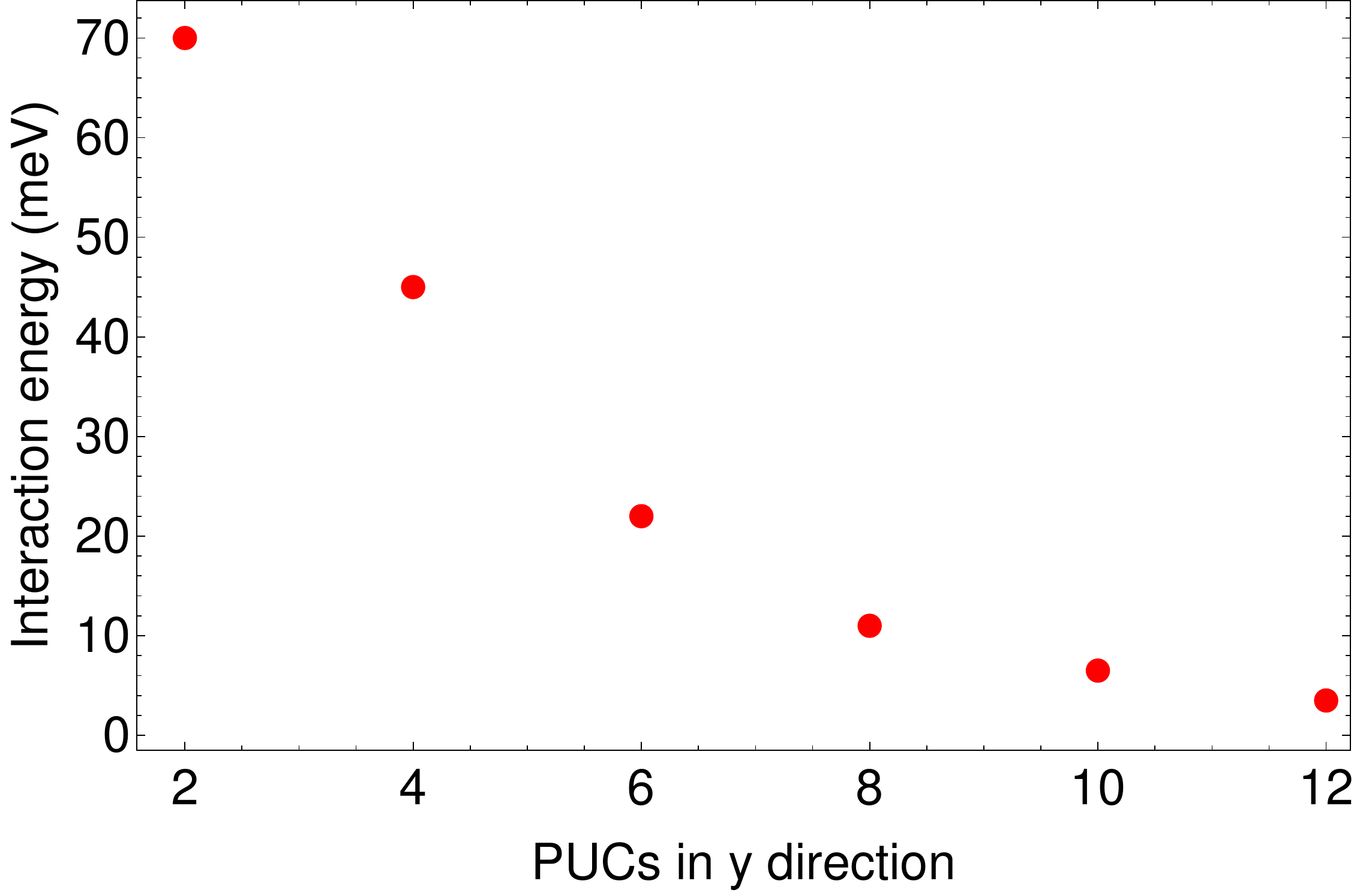}
        \caption{$y$ direction.}
    \end{subfigure}
    \caption{Intra-cellular interaction energy between adjacent N defects as a function of lateral slab dimensions.}\label{fig:int}
\end{figure}

In Figure~\ref{fig:int} we plot the interaction energy $\beta$ as a function of the number of primitive unit cells in the $x$ and $y$ directions. Slabs constructed to demonstrate the dependence on $x$ were two PUCs wide in the $y$ direction, while those constructed to demonstrate the dependence of $y$ were two PUCs wide in the $x$ direction. This was done to avoid direct bonding between adjacent N defects. As expected, the interaction energy decreases with increasing spatial separation. Thinner slabs display interaction energies of up to 120~meV, while all interactions become negligible for slabs six PUCs wide in either the $x$ or $y$ directions.

\subsection{C insertion into C--N}

\begin{figure}
    \centering
    \begin{subfigure}[b]{0.4\textwidth}
        \includegraphics[width=\textwidth]{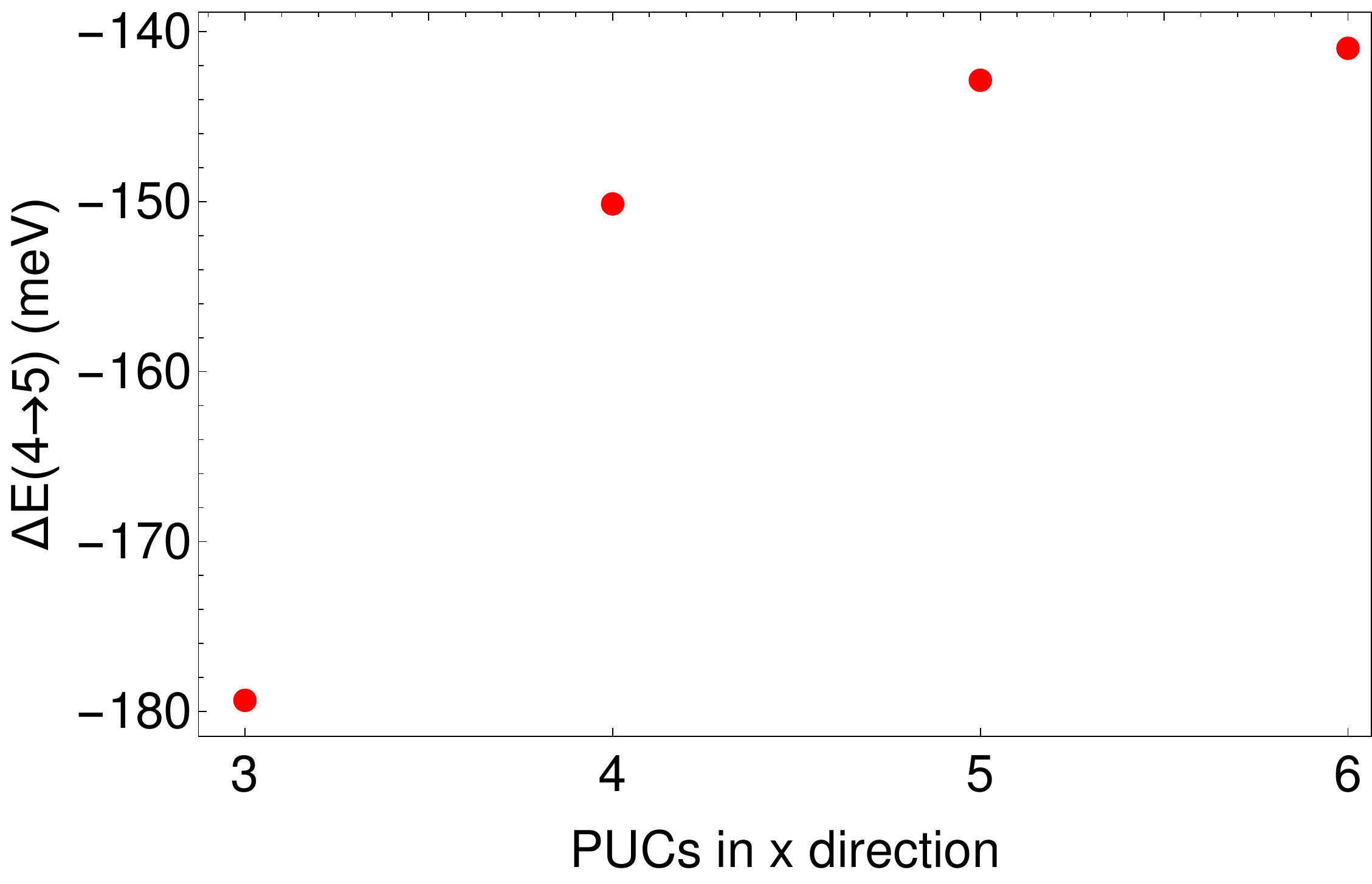}
        \caption{$\Delta E(4\rightarrow5)$.}
    \end{subfigure}
    ~ 
    \begin{subfigure}[b]{0.4\textwidth}
        \includegraphics[width=\textwidth]{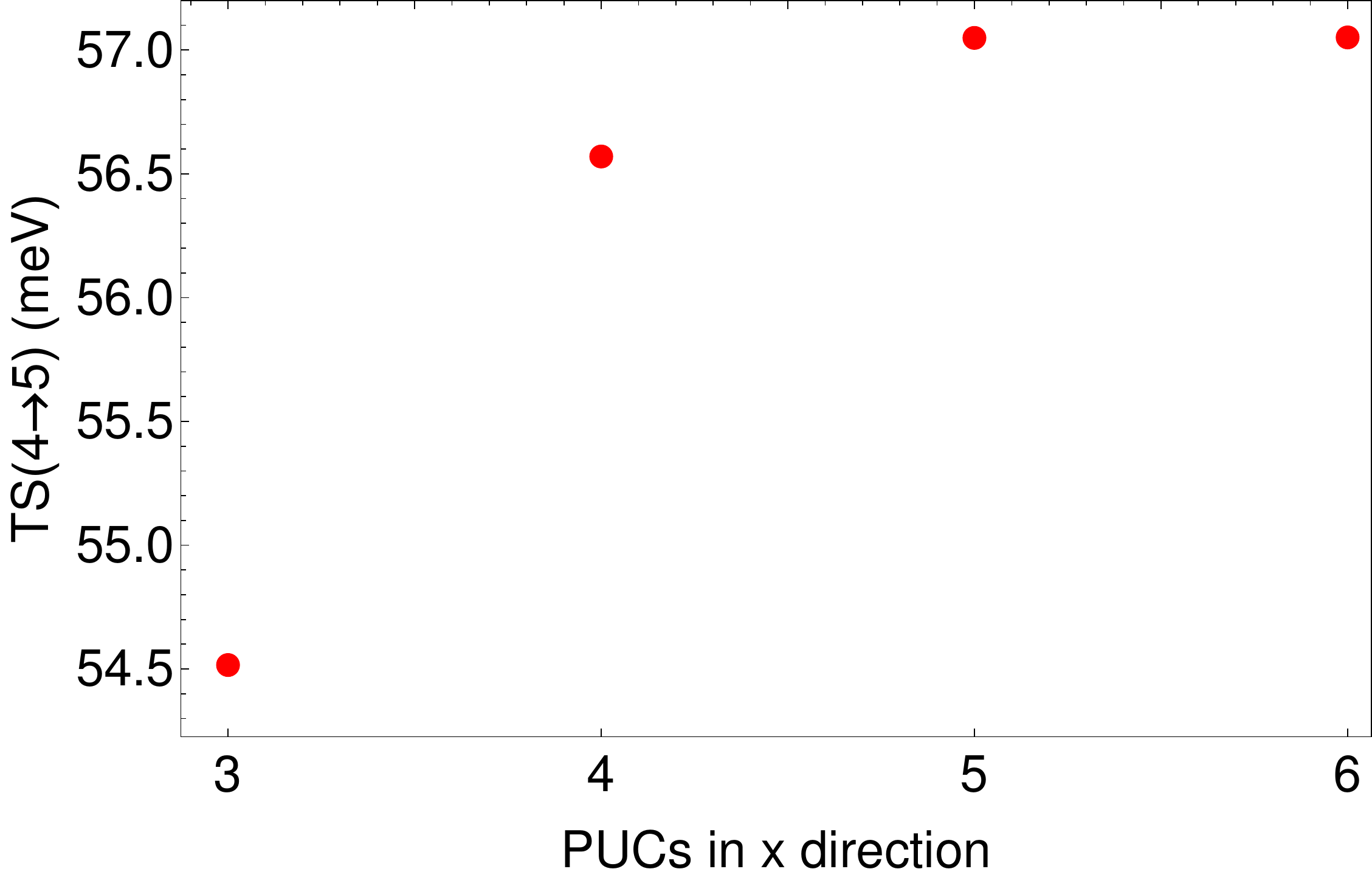}
        \caption{TS$(4\rightarrow5)$.}
    \end{subfigure}
    
    \begin{subfigure}[b]{0.4\textwidth}
        \includegraphics[width=\textwidth]{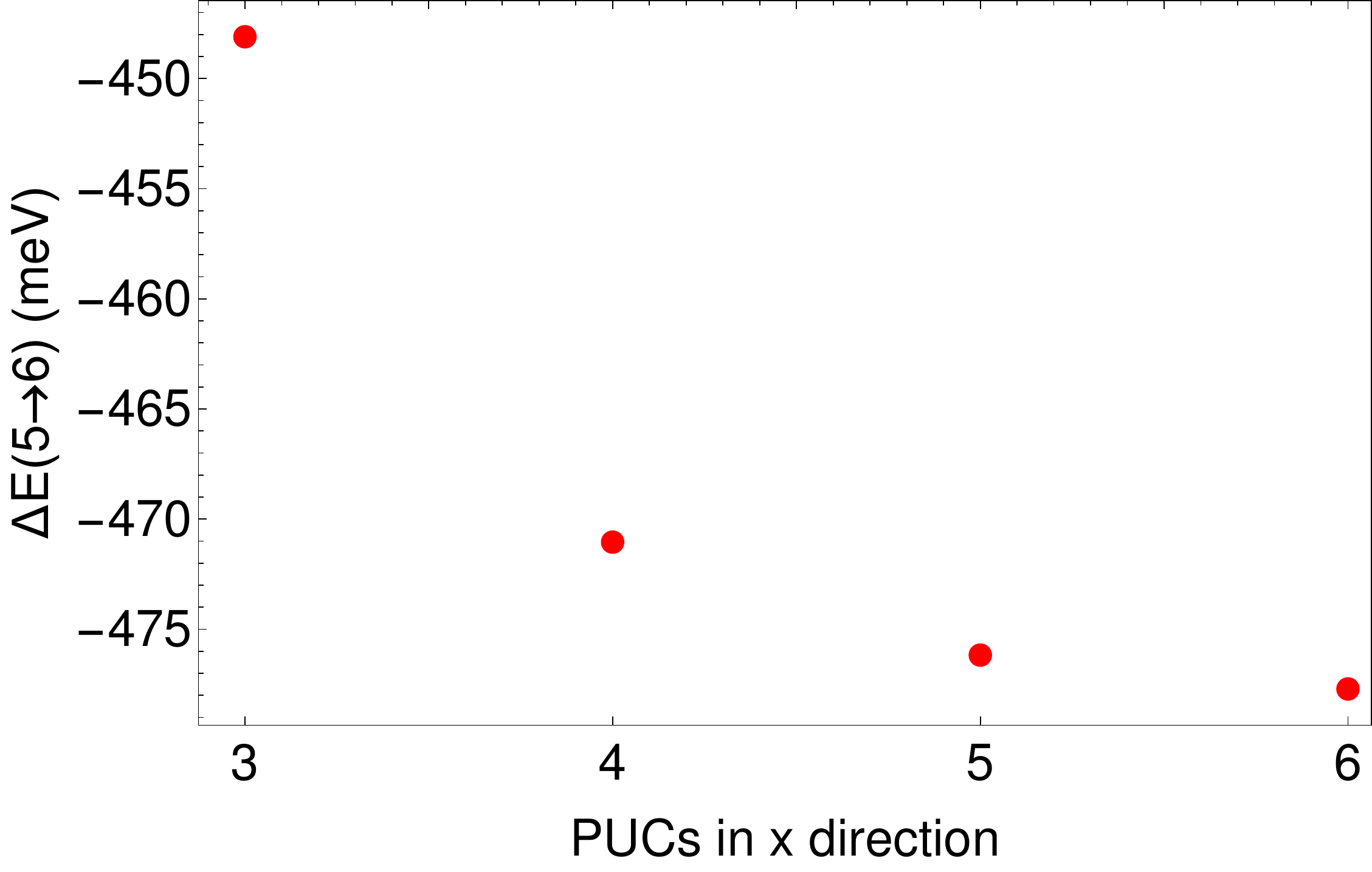}
        \caption{$\Delta E(5\rightarrow6)$.}
    \end{subfigure}
    ~
        \begin{subfigure}[b]{0.4\textwidth}
        \includegraphics[width=\textwidth]{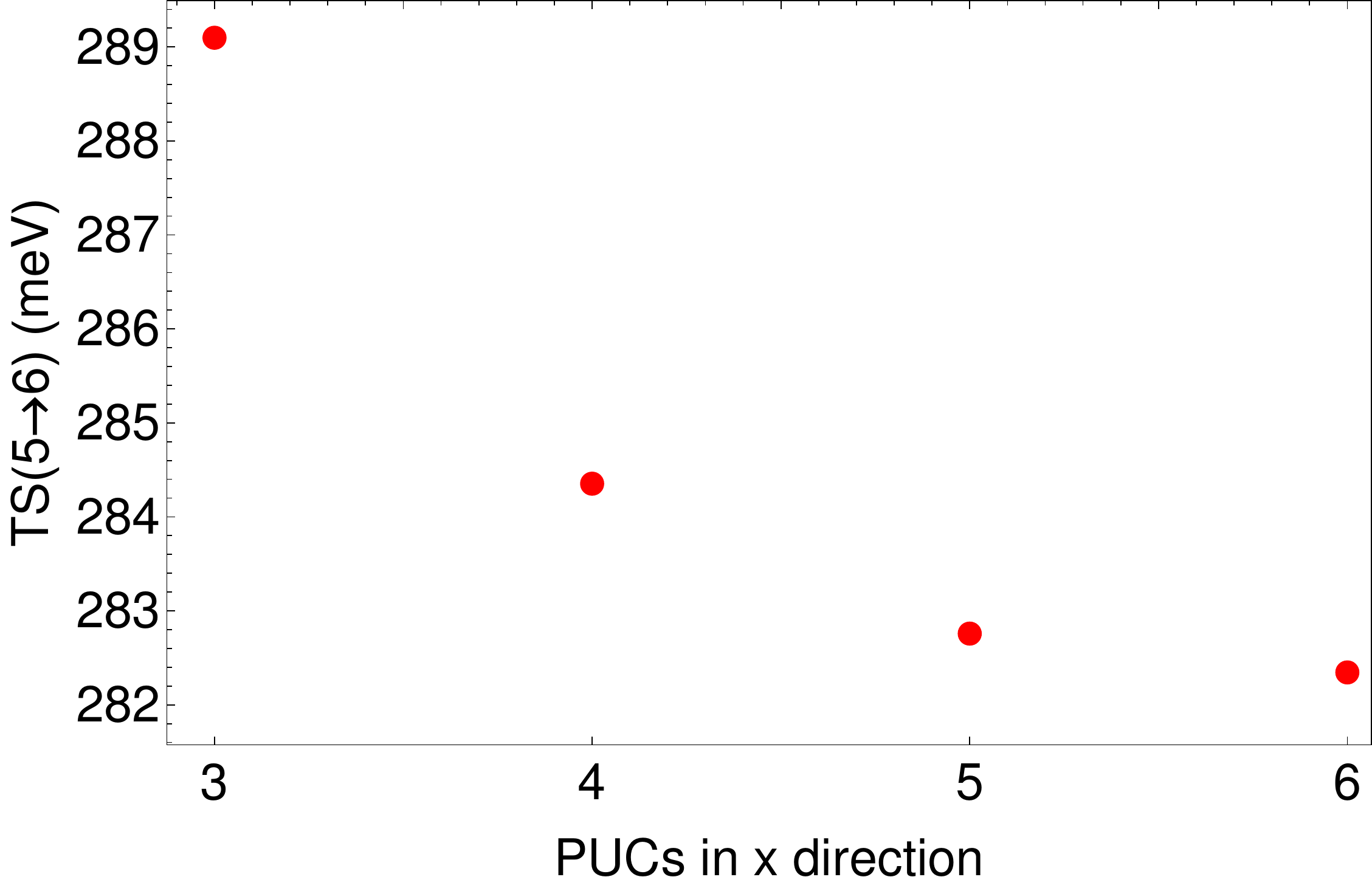}
        \caption{TS$(5\rightarrow6)$.}
    \end{subfigure}
    \caption{Energetics of steps $4\rightarrow5$ and $5\rightarrow6$ for C insertion into a C--N dimer as a function of slab width in the $x$ direction.}\label{fig:XLat}
\end{figure}

\begin{figure}
    \centering
    \begin{subfigure}[b]{0.4\textwidth}
        \includegraphics[width=\textwidth]{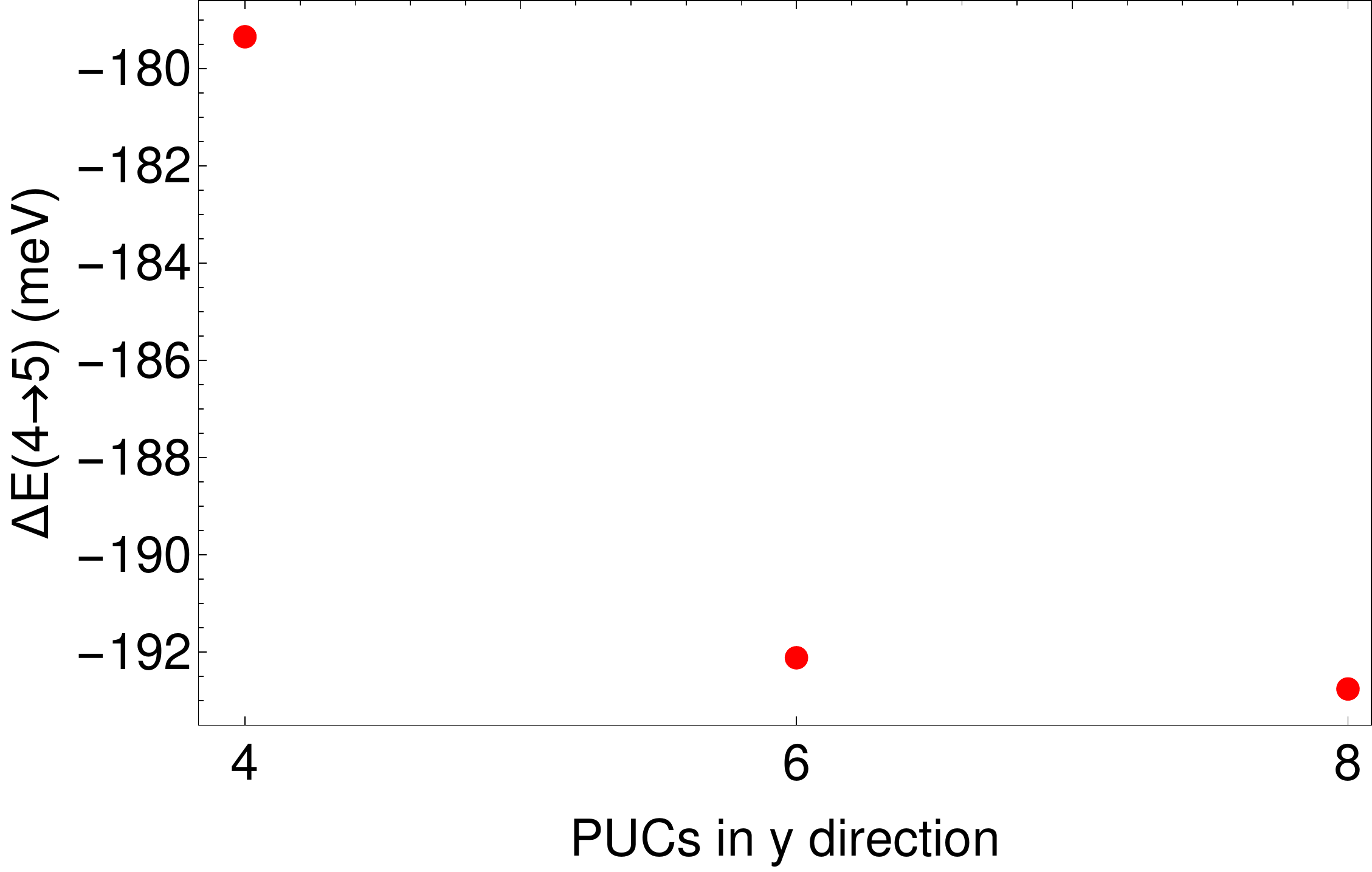}
        \caption{$\Delta E(4\rightarrow5)$.}
    \end{subfigure}
    ~ 
    \begin{subfigure}[b]{0.4\textwidth}
        \includegraphics[width=\textwidth]{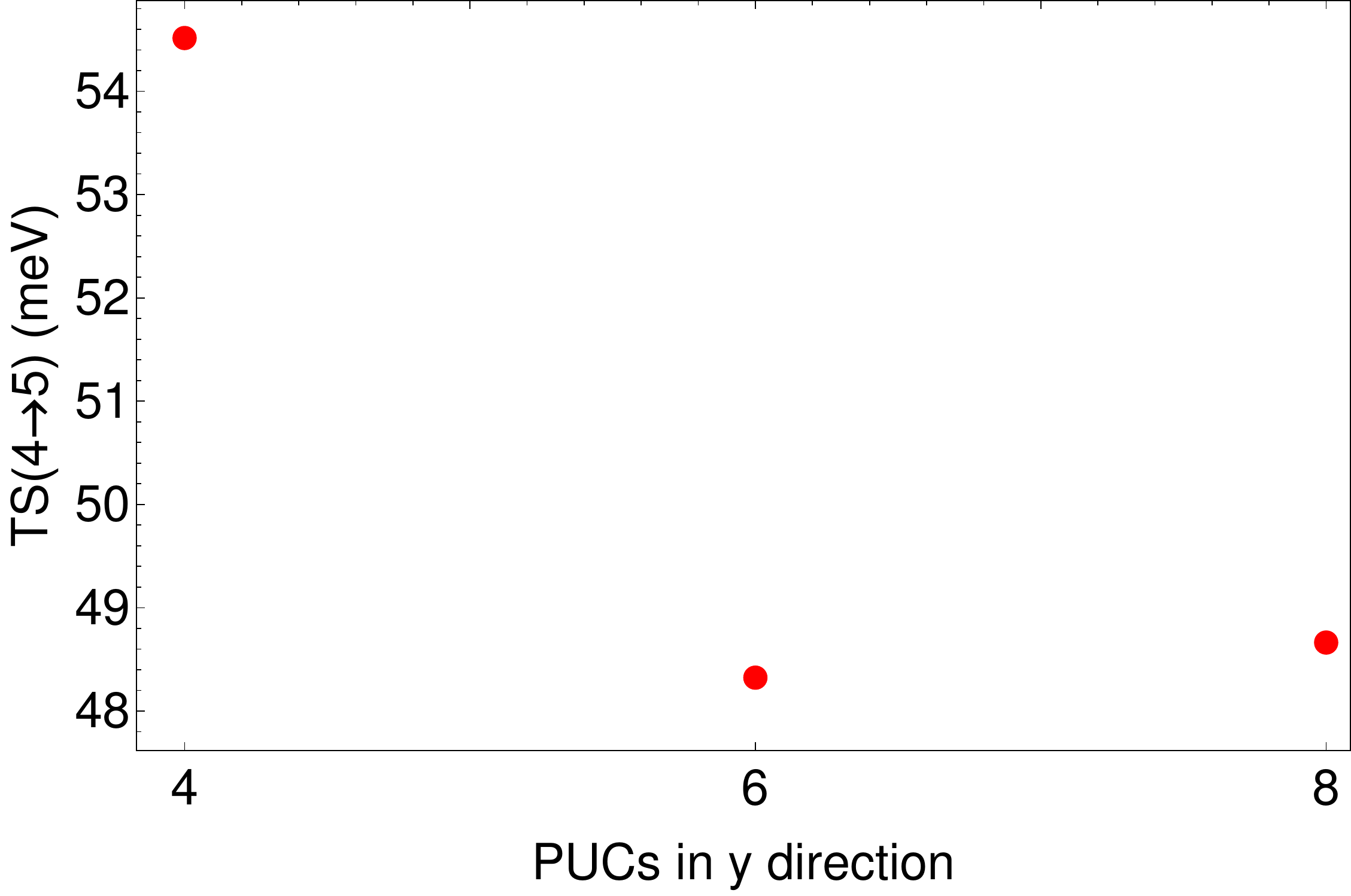}
        \caption{TS$(4\rightarrow5)$.}
    \end{subfigure}
    
    \begin{subfigure}[b]{0.4\textwidth}
        \includegraphics[width=\textwidth]{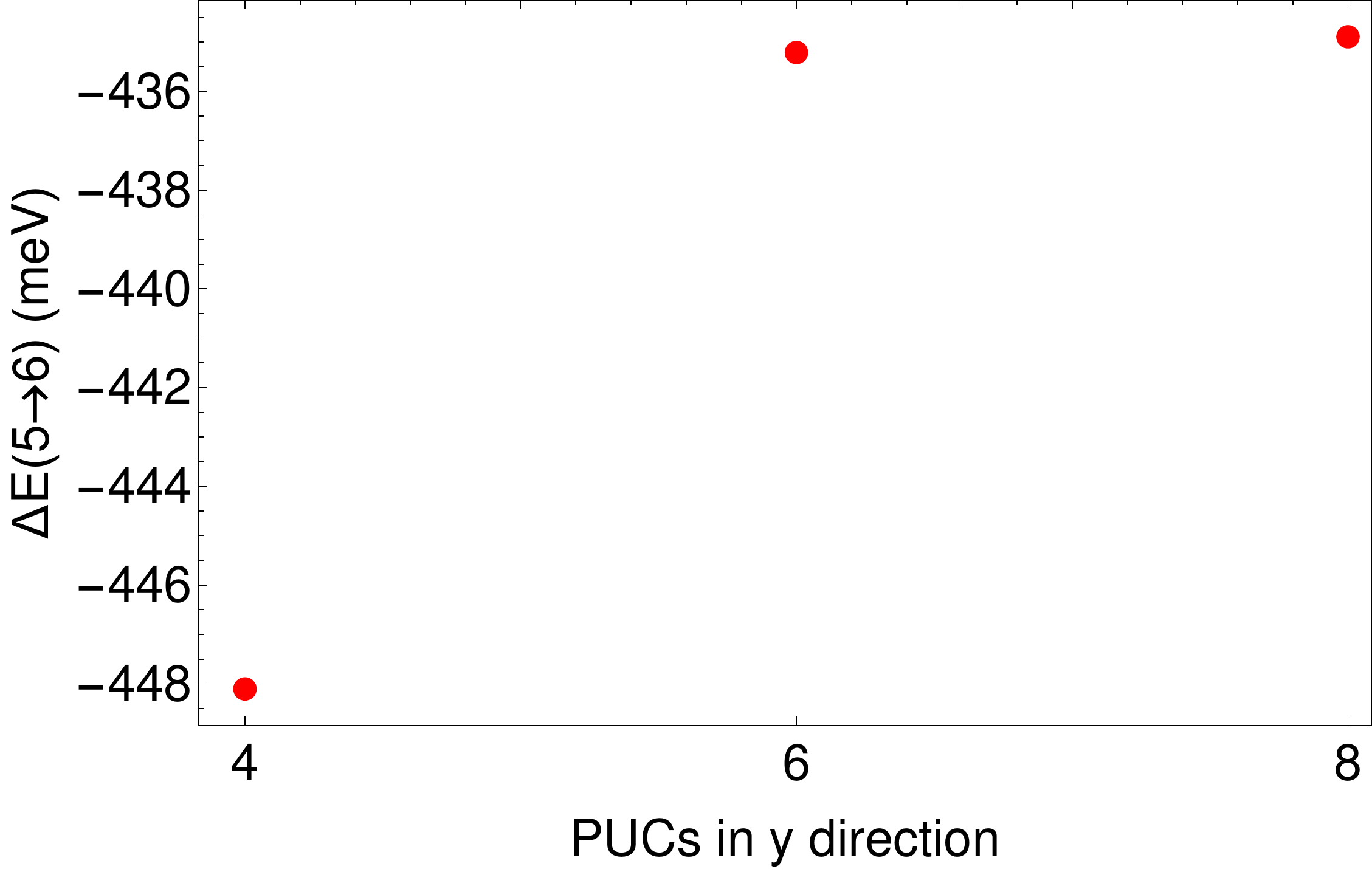}
        \caption{$\Delta E(5\rightarrow6)$.}
    \end{subfigure}
    ~
        \begin{subfigure}[b]{0.4\textwidth}
        \includegraphics[width=\textwidth]{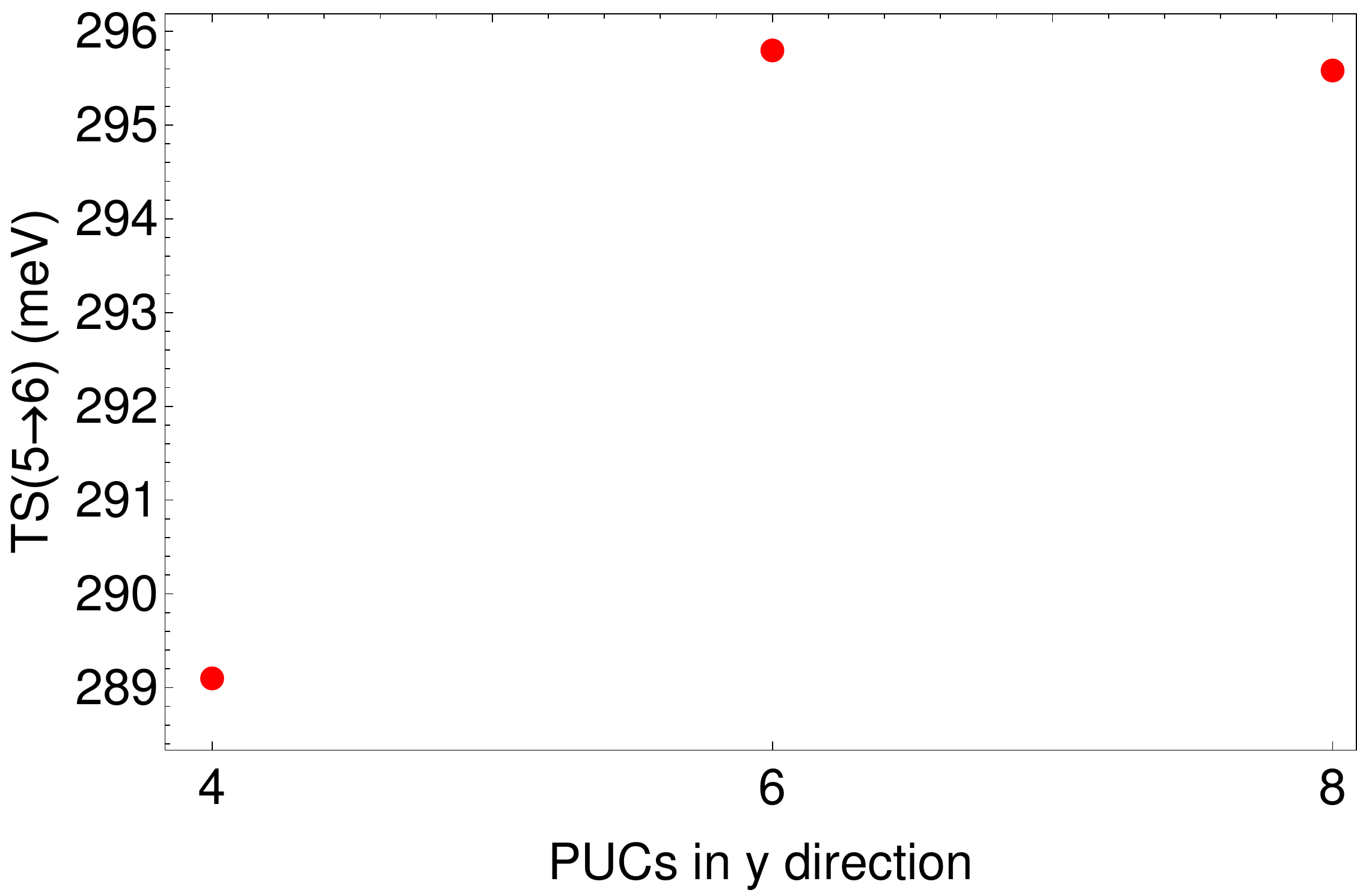}
        \caption{TS$(5\rightarrow6)$.}
    \end{subfigure}
    \caption{Energetics of steps $4\rightarrow5$ and $5\rightarrow6$ for C insertion into a C--N dimer as a function of slab width in the $y$ direction.}\label{fig:YLat}
\end{figure}

In Figures~\ref{fig:XLat} and \ref{fig:YLat} we present the relationship between $x$ and $y$ slab dimensions and the energetics of steps $4\rightarrow5$ and $5\rightarrow6$ for C insertion into the C--N dimer. As previously, slabs constructed to demonstrate the dependence on $x$ were two PUCs wide in the $y$ direction, while those constructed to demonstrate the dependence of $y$ were two PUCs wide in the $x$ direction.

Figure~\ref{fig:XLat} reveals that reaction energies have the greatest dependence on number of PUCs in the $x$ direction, differing by up to 40~meV for $\Delta E(4\rightarrow5)$ and 25~meV for $\Delta E(5\rightarrow6)$ between cells three and six PUCs wide. The dependence of slab width on transition energy is much less pronounced, differing by a maximum of 7~meV. Similar trends are observed in Figure~\ref{fig:YLat} for the dependence on PUCs in the $y$ direction. Note that these increase in multiples of two due to the diamond surface reconstruction. We find that $\Delta E(4\rightarrow5)$ and $\Delta E(5\rightarrow6)$ differ by up to 12~meV for slabs between four and eight PUCs thick, with the corresponding transition state energies differing by only 7~meV. Despite our limited data for dependence on PUCs in the $y$ dimension, it appears that all energies converge beyond eight PUCs 
in agreement with the results seen for the $x$ dependence.

The energy difference observed in the reaction energetics as a function of slab dimensions is considerably lower than that observed in the interaction energy of N defects. This is attributable to the fact that N is embedded within the lattice and is therefore capable of long-range interactions through bonding with three adjacent C ions. This is not the case for C insertion, in which the methyl group is adsorbed to the surface and so has a less pronounced effect on overall surface geometry.

Based on these results we choose for our final slab geometry five PUCs in the $x$ direction and four in the $y$ direction, as depicted in Figure~\ref{fig:surf}. Five PUCs in the $x$ direction result in reaction barriers and energies within 1--2~meV of the converged result, and so we sacrifice no major losses in accuracy with this choice. Based on the results in Figure~\ref{fig:YLat} it would be desirable to choose a slab at least six PUC wide in the $y$ direction. However, this is beyond our computational capabilities for the B3LYP functional and so we have compromised an accuracy of at most 12~meV by selecting the four PUC wide slab.

\section{Instability of state 5 during C insertion into C--C}
\label{sec:instab}

\begin{figure}[]
    \centering
    \includegraphics[width=0.8\textwidth]{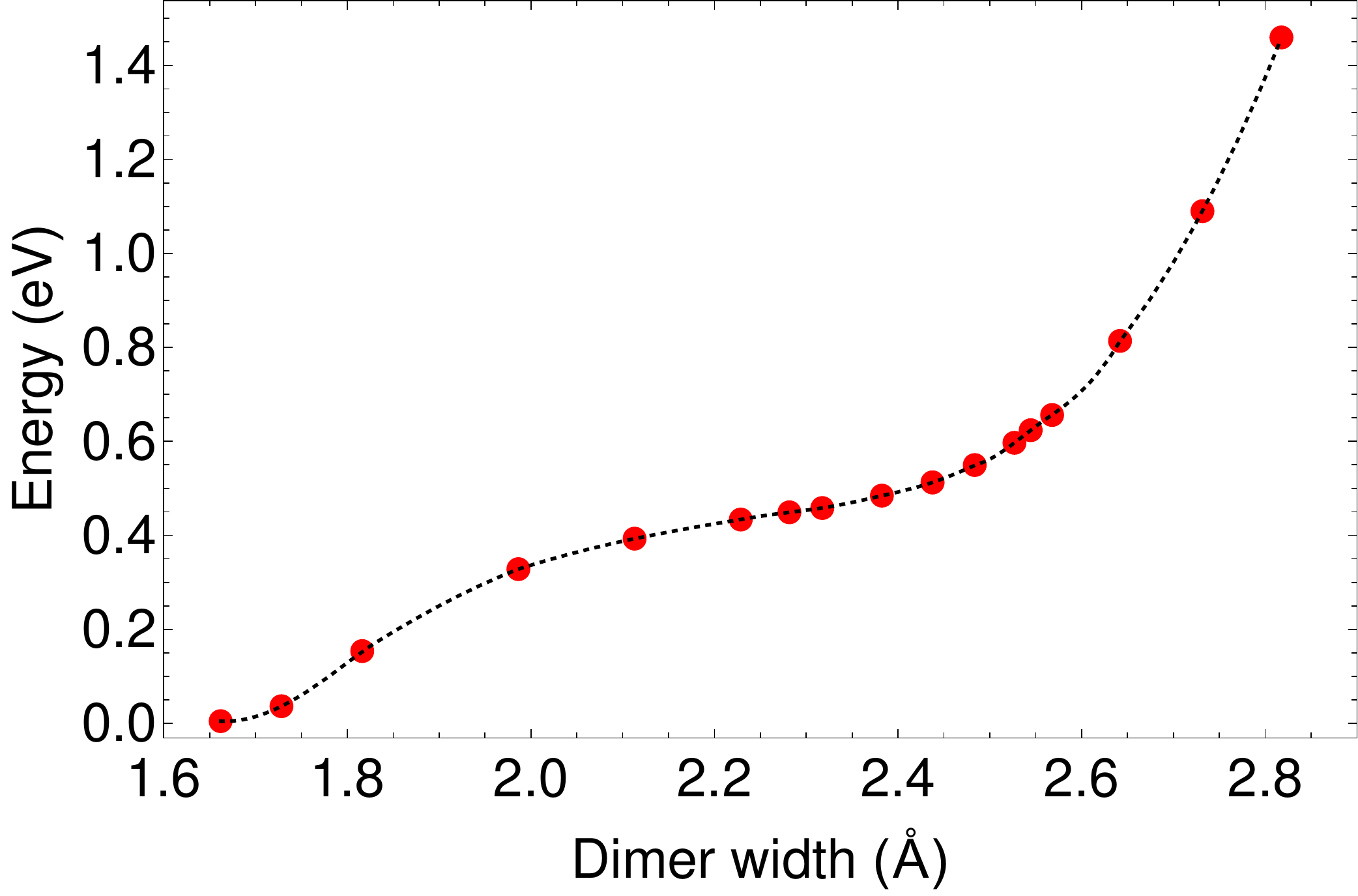}
    \caption{Slab supercell energy as a function of C--C dimer width using PBE functionals. This represents the change in energy to the system during the ring opening mechanism ($4\rightarrow5$). The dimer width of $\approx0.17$~\AA \ corresponds to state 4 in Figure~1 of the main text. No minimum is observed, indicating that state 5 of the C insertion mechanism is a transition state instead of a stable geometry. The dotted line is a guide for the eye.}
    \label{fig:dimer_open}
\end{figure}

The instability of state 5 during C insertion into the C--C dimer was further investigated by determining the relationship between slab energy and dimer width using the PBE functional. To do so we fix the lateral position of the C which bonds with the CH${}_2$ adsorbate while allowing all other ions to relax. If state 5 is stable, we should expect to see a local energy minimum when the dimer bond is dissociated, corresponding to a C--C bond length of $\approx$2.4~\AA. However, as shown in Figure~\ref{fig:dimer_open}, we observe that the dissociated dimer resides on a low gradient of the potential energy surface. Consequently, our calculations indicate that state 5 is a transition state.

As presented in the main text, our adamantane cluster model identifies that state 5 is stable with a dimer separation distance of 2.95 \AA. The results in Figure~\ref{fig:dimer_open} indicate that realising this separation distance in the slab model would require overcoming an activation barrier of at least 1.4~eV, which is highly improbable at typical growth temperatures. Hence, this indicates that cluster models may produce over-relaxed and unphysical geometries.

Consequently, our slab results demonstrate that the direct transition $4\rightarrow6$ is required to achieve C insertion into a C--C dimer. Interestingly, the studies by Cheesman\textit{et al.}\cite{Cheesman2006} and Kang and Musgrave\cite{Kang2000} considered this transition and both identified activation energies in excess of 2~eV. Our considerably lower estimates of 0.96~eV presented in the main text may be attributed to the climbing NEB technique, which performs a rigorous search of the potential energy surface.

\newpage

\bibliography{supplementary}